%% file: paper.tex
\documentclass[12pt, reqno]{amsart}

\usepackage[thm_section]{macros}
\newgeometry{margin=1in}

\usepackage[all=normal, paragraphs=tight, bibbreaks=tight, floats=tight,bibnotes=tight]{savetrees}

\usepackage{times} 

\title{The purpose of an estimator is what it does:\\ Misspecification, estimands, and over-identification}
\author{Isaiah Andrews, Jiafeng Chen, and Otávio Tecchio\footnote{Andrews: Department of Economics,
Massachusetts Institute of Technology, and NBER, iandrews@mit.edu; Chen: Department of Economics, Stanford  University, jiafeng@stanford.edu; Tecchio: Department of Economics, Massachusetts Institute of Technology, otecchio@mit.edu.  We thank Anh Nguyen and Alissa Kopylova for excellent research assistance, and Hiroaki Kaido, Patrick Kline, Francesca Molinari, Jesse Shapiro, an anonymous referee, and participants at the 2025 Econometric Society World Congress session on model misspecification for helpful comments.}}

\date{\today}

\usepackage{amsthm}

\usepackage{booktabs}
\usepackage[flushleft]{threeparttable}

\newcommand{\avar}{\text{aVar}}
\newcommand{\plim}{\text{plim}}

\begin{document}
\maketitle

\begin{abstract} 
In over-identified models, misspecification---the norm rather than
 exception---funda-\\mentally changes what estimators estimate. Different estimators imply
 different estimands rather than different efficiency for the same target. A review of
 recent applications of generalized method of moments in the \emph{American Economic
 Review} suggests widespread acceptance of this fact: There is little formal
 specification testing and widespread use of estimators that would be inefficient were
 the model correct, including the use of ``hand-selected'' moments and weighting
 matrices. Motivated by these observations, we review and synthesize recent results on
 estimation under model misspecification, providing guidelines for transparent and robust
 empirical research. We also provide a new theoretical result, showing that Hansen's
 J-statistic measures, asymptotically, the range of estimates achievable at a given
 standard error. Given the widespread use of inefficient estimators and the resulting
 researcher degrees of freedom, we thus particularly recommend the broader reporting of
 J-statistics.
\end{abstract}

\newpage

\section{Introduction}

Scientific and policy questions in economics frequently hinge on quantities  such as
counterfactual outcomes or effects on welfare. Recovering them often requires strong
modeling assumptions.  At the same time, many economists concede that economic models are
at best approximately true \citep{white1982maximum, hansen2008robustness,
manski2013policy,rodrik2015econ}. This leads to a gap between formal econometric
recommendations and empirical practice for over-identified models.\footnote{That is,
models which admit different estimators for the same parameter and consequently imply
testable restrictions on the data distribution, c.f. \citealt{chen2018overidentification}.}

Many textbook treatments of over-identified models \citep[e.g.][]{hyashi2000econometrics, stock2018econometrics, davidson2003econometrics} recommend
researchers test whether the model's over-identifying restrictions hold and use an
efficient estimator when failing to reject. Experience reading economics papers and
attending economics seminars, however, suggests  that empirical practice often flouts these
prescriptions. To document  this, we examined empirical papers published in the 
\emph{American Economic Review} between 2020 and 2024 which fit models using the
 generalized method of moments \citep[GMM,][]{hansen1982gmm} or related methods,
 obtaining a sample of 36 papers, of which 22 are over-identified.  In this sample, we
 find that researchers commonly fit models using only a subset of the model's
 implications, for instance fitting a fully parametric model by moment-matching rather
 than maximum likelihood. 
 Researchers also commonly weight moments differently than would the optimal estimator (e.g. 14 of the over-identified papers in our sample report no efficiently-weighted specifications).
 Both practices yield inefficient estimators under correct model specification.  Formal
 tests of over-identifying restrictions are notably scarce (only 3 papers in our sample
 report J-tests or J-statistics of overidentifying restrictions), though various ``eyeball tests'' are
 common, with researchers e.g. plotting the gap between data- and model-implied moments (25 papers in our sample).

This gap between econometrics as taught and econometrics as practiced could appear
puzzling if one thought economic models were plausibly correct---after all, it would be
better to use a correct model, and to use the best estimator under that model, than to
persist in using an incorrect model or an inefficient estimator.  The common practices
outlined above are, on the other hand, reasonable under the premise that any model we can
take to the data is almost certainly wrong.  Consistent with this premise, many papers in
our data show evidence of substantial misspecification, with 8 papers reporting \emph
{just-identified} specifications where the parameter estimates cannot set the moments to
zero.  Under this view, it is natural to  fit the model aiming for the ``least-bad''
answer. In judging model fit, it is moreover natural to draw quantitative judgments about
where and how well the model fits, rather than focusing on the all-or-nothing question of
whether the model fits exactly.

A similar perspective underlies a large and active recent literature in econometrics which
studies the behavior and interpretation of popular estimators without assuming that the
model is correct.  This article provides a selective synthesis of  this literature.
While one can reasonably critique some of this literature as a reverse-engineering
exercise---starting from estimators and working backwards in the hope of finding
interpretable estimands \citep{mogstad2024IV}---the perspective of this
literature can also reasonably be summarized as ``the purpose of an estimator is what it
does.''\footnote{Paraphrased from \cite{beer1985diagnosing}---%
``The purpose of a system is what it does.  There is, after all, no point in claiming that
the purpose of a system is to do what it consistently fails to do.''}  That is, given
that economic models are generally misspecified, it is important to understand how
estimators actually behave in applications. Moreover, in our view, it can be reasonable to choose among estimators on this basis.

While reasonable, the common practice of choosing estimators based on misspecification concerns increases researcher degrees of freedom \citep{Simmonsetal2011}.  Different researchers estimating the same economic model on the same data may reach different conclusions due to differing  misspecification concerns and subsequent estimation choices.  Worse,  dishonest researchers could use misspecification concern as a pretext for estimation choices which deliver a desired result, e.g. rejection of some null hypothesis. 

As a first step towards quantifying the scope for such manipulation, and motivated by the wide range of weight-selection approaches observed in our survey of AER papers, we consider the choice of weight matrices for a given (over-identifying) set of moments.  In models where the degree of misspecification is on the same order as sampling uncertainty (i.e. locally misspecified models) we prove that, asymptotically, the set of estimates obtainable at a given standard error is an interval centered at the efficient GMM estimate, with width proportional to the square root of \cite{hansen1982gmm}'s J-statistic.  Moreover, we show that a nefarious researcher can choose weights to engineer a t-statistic value of at least $\sqrt{J}$ for any null hypothesis.  Thus, given a set of moments, J-statistics measure the scope for a nefarious researcher to ``weight hack'' their findings.

To summarize our conclusions, we recommend the following:
\begin{enumerate}
\item Researchers should distinguish concerns and diagnostics for correct \emph{econometric}
 specification (i.e. does the model correctly describe the relationship between the
 observable data and the objects of economic interest) vs. correct \emph{statistical}
 specification (i.e. does the model fit the data distribution).

\item Researchers should be explicit about what their choice of estimator implies for
 their estimand.  If they use estimators which are inefficient under the model (e.g. due
 to non-optimal weighting), they should articulate why they believe such estimators are
 preferable.

\item Researchers should use misspecification-robust standard errors, which excludes the
 usual  GMM standard errors \citep{hansen1982gmm} in over-identified models.  Formulas are
 available for many problems, and when in doubt the bootstrap 
will often work.

\item Researchers estimating over-identified moment condition models should report
 J-statistics, since they summarize the range of results that may be obtained via
 alternative weighting choices. J-statistics are thus important to report even if one dispenses with
 J-tests.
\end{enumerate}

The following sections develop these points in detail, highlighting relevant references
from the literature as we proceed.  \Cref{sec: Misspecification} defines what we mean by
econometric and statistical models, illustrating with examples, and introduces corresponding
notions of misspecification.  \Cref{sec: estimands} defines the estimands resulting from
given estimators, and reviews some important estimand characterization results.  \Cref
{sec: inference on estimands} discusses misspecification-robust standard errors.
Finally, \cref{sec: local misspec} reviews the idea of local misspecification and
derives what is to the best of our knowledge a new result, relating J-statistics to the
range of estimates attainable under a standard error bound.  \Cref{sec: Limit problem}
details a normal limit problem used to prove our J-statistic result, while \cref
{sec: proofs} collects proofs for results stated in the main text.

\section{What Do We Mean by ``Misspecified?''}\label{sec: Misspecification}

Suppose we observe a sample of $n$ observations $X^n=\{X_1,...,X_n\}$ drawn from a distribution $P$.\footnote{Formally, we take $P$ to be a distribution
over the infinite sequence $X^\infty=\{X_1,X_2,...\}$ and suppose the sample is the first
$n$ elements.  Thus, our analysis covers both settings with iid data (in which case $P$ is
a product measure) and cases with dependent data, subject to restrictions we note below.}
We are interested in using the data to learn an \emph{economic parameter}
$\theta\in\Theta,$ such as a counterfactual outcome or optimal policy.
For the data to convey information about $\theta$, we must restrict which parameters can accompany a given data distribution.
These restrictions could come from economic theory, from empirical evidence in other
settings, or from ad-hoc restrictions chosen for analytical convenience.  These
restrictions imply an \emph{econometric model} 
$\mathcal{M}\subseteq\Theta\times\Delta(\mathcal{X}),$
for $\Delta(\mathcal{X})$ the set of distributions on the sample space $\mathcal{X}.$ The
model $\mathcal M$ collects pairs of $(\theta, P)$ that are consistent with our
restrictions. $\mathcal M$ fully describes the relationship between the data distribution
and the quantity of interest, and so determines the identification status of the
parameter.\footnote{While we discuss $\mathcal{M}$ as an econometric model given our focus on the economics literature, one might prefer another term, e.g. substantive model, since the same points arise when studying causal or counterfactual questions outside of economics.   \cite{spanos2016transforming} discusses related points under the heading of substantive adequacy.}  Specifically, the identified set for $\theta$ under data distribution $P$
is the set of parameter values consistent with $P$:
$
\Theta_I(P)=\left\{\theta\in\Theta:(\theta,P)\in\mathcal{M}\right\}. $ $\theta$ is
 point-identified if and only if $\Theta_I(P)$ is a singleton for all $P$ consistent with
 $\mathcal{M}$.

We call the set of all $P$ consistent with $\mathcal{M}$ the \emph{statistical model}: 
\[
\mathcal{P}=\left\{P\in\Delta(\mathcal{X}):\text{ there exists } \theta\in\Theta \text{ such that } (\theta,P)\in\mathcal{M}\right\}.
\]
Provided the parameter $\theta$ is point-identified under $\mathcal{M}$, we can define a \emph{statistical
parameter} $\theta^*:\mathcal{P}\to\Theta$ such that $\theta^*(P)=\theta$ for all $(\theta,P)\in\mathcal{M}$. By construction, $\theta^*$ is
point-identified as well: $\Theta_I (P)= \{\theta^* (P)\}$
for all $P\in \mathcal{P}$.

\begin{exsq}[Linear Causal Model] Suppose we observe $X_i=(Y_i,W_i)$ drawn iid.  Here
$Y_i\in\mathbb{R}$ is a scalar outcome while $W_i\in\mathcal{W}\subseteq\mathbb{R}$ is a continuous treatment.
We  believe that (i) each individual has potential outcome $Y_i(\cdot),$ where the
observed outcome for individual $i$ is equal to that individual's potential outcome
evaluated at realized treatment, $Y_i=Y_i(W_i)$; that (ii) treatment is randomly assigned
with $W_i$ independent of $Y_i(\cdot)$; and that (iii) the potential outcomes are linear
in $w,$
\[
Y_i(w)=\alpha_i+\beta_i\cdot w.
\] This is a special case of the random coefficient models studied by e.g. \citet
 {beran1992estimating}.\footnote{Note that if we take $W_i\in\{0,1\},$ rather than continuous, then this example is simply the potential outcomes model, $\alpha_i=Y_i(0)$ and $\beta_i=Y_i(1)-Y_i(0).$} %
Our parameter of interest is the average derivative $\theta = \E\left[\frac{\partial}{\partial
w} Y_i(W_i)\right],$ which describes the average effect of a marginal increase in $W_i.$  Under the model this average derivative can be written as $\E[\beta_i].$

Under an additional regularity condition on a variance, which we shorthand with
$\Xi = \var(W_i, \alpha_i, \beta_i)$, our econometric model is
\[
\mathcal{M}=\left\{\pr{\E[\beta],\bigtimes_{i=1}^\infty\mathcal{L}
(\alpha+\beta W,W)}: (\alpha,\beta)\indep W,\det(\Xi)\in(0,\infty)\right\}
\]
where we use $\mathcal{L}(A_i)$ to denote the law of a random variable $A_i.$  The statistical model collects the corresponding data distributions, 
\[
\mathcal{P}=\left\{\bigtimes_{i=1}^\infty\mathcal{L}(\alpha+\beta W,W):
(\alpha,\beta)\indep W,\det
(\Xi)\in(0,\infty) \right\}.
\]
If the econometric model is correct, least squares regression of $Y_i$ on $W_i$ recovers
$\theta,$ so we can define the statistical parameter as $\theta^*(P)=\frac{\cov_P
(W_i,Y_i)}{\var_P(W_i)}.$   
\end{exsq}

\begin{exsq}[Generalized Method of Moments]

Suppose the distribution of $(X_1,\ldots,X_n)$ under $P$ is stationary. Suppose the
parameter of interest $\theta$ is, under the model, some function of a full set of model
parameters $\psi
\in\Psi\subseteq\mathbb{R}^p $, $\theta=\vartheta(\psi)$ for a known function $\vartheta
 (\cdot)$. Our model implies that a vector of moment conditions \[\E_P[g
 (X_i,\psi)]=0\in\mathbb{R}^k\] holds at the true value of $\psi.$  

Let the econometric model $\mathcal{M}$ denote the set $(\theta,P)$ consistent with the
moment restrictions, along with additional regularity conditions (e.g., finite moments, stationarity)
encoded by a set $ \mathcal{P}^R:$
\[
\mathcal{M}=\left\{(\vartheta(\psi),P):\psi\in\Psi,P\in\mathcal{P}^R, \E_P[g
 (X_i,\psi)]=0\right\}. \numberthis 
 \label{eq:gmm_model}
\]
This econometric model implies a statistical model
\[\mathcal{P}=\left\{P\in\mathcal{P}^R:\text{ there exists }\psi\in\Psi\text{ such
 that }\E_P[g(X_i,\psi)]=0\right\}.\] Provided the identified set for $\psi$ is a
 singleton under each $P\in\mathcal{P}$---which in this context means that the moment
 equation $\E_P[g(X_i,\psi)]=0$ is solved at a unique $\psi$---we can define $\psi^*:\mathcal{P}\to\Psi$ implicitly as the solution to the moment equations, $\E_P[g
 (X_i,\psi)]=0,$ and can define the statistical parameter as $\theta^*(P)=\vartheta
 (\psi^*(P)).$
\end{exsq}

So long as (a) we are  confident in the specification of our model and (b) the parameter $\theta$
is point-identified under $\mathcal{M}$, the distinction between the econometric and
statistical models is inconsequential, since the economic parameter $\theta$ and the
statistical parameter $\theta^*(P)$ always agree.  Once we consider misspecification,
however, this distinction starts to matter.  

We say that the model is \emph{econometrically misspecified} if the true $ (\theta,P)$ pair
falls outside the econometric model, $(\theta,P)\not\in\mathcal {M}.$ We assume the parameter
$\theta$ remains well-defined under misspecification, which restricts the sorts of
parameters we consider.  In particular, our analysis is well-suited to, e.g., average
counterfactual outcomes or causal effects, which may be defined (if not always estimated)
non-parametrically.  By contrast, for parameters which are ill-defined outside a specific
model (e.g., preference parameters tied to a specific functional form), it makes little sense to contemplate a ``true'' $\theta$ outside of the model. Fortunately, in many economic analyses the objects
of ultimate interest (e.g., counterfactuals, welfare, and optimal policies) are
well-defined under much weaker assumptions than those often used to conduct estimation.

Analogously, the model is \emph{statistically misspecified} if the observed data
distribution rejects it, $P\not\in\mathcal{P}.$  Since we have thus
far defined the statistical parameter $\theta^*$ only on the model, $\theta^*(P)$ is
undefined under statistical misspecification.   Statistical misspecification implies
econometric misspecification but not the reverse.

\begin{exsq}[Generalized Method of Moments, continued]
    
The moment condition model is econometrically misspecified at $(\theta,P)$ if and only if the
moment conditions are violated at all values of $\psi$ consistent with the true economic
parameter $\theta,$ i.e.
\[
0\not\in \left\{E_P[g(X_i,\psi)]:\psi\in\Psi,\vartheta(\psi)=\theta\right\}.
\]
For instance, it might be that the moment conditions are not solved by any parameter
value  $\psi$ consistent, through $\vartheta$, with the true counterfactual outcome $\theta$.
Under this definition, the reason for misspecification $(\theta, P) \not\in \mathcal
M$---incorrect relationship between $\psi$ and $P$  (i.e. the moments $g$), incorrect
relationship between $\psi$ and $\theta$ (i.e. the function $\vartheta$), or some other
source---is irrelevant. Indeed, since correct specification requires only that our model $
\mathcal{M}$ allow the true $(\theta,P)$ pair, it is possible for the model to be
correctly specified even if our underlying economic assumptions are incorrect and the
model parameter $\psi$ does not have a well-defined ``true'' value. Similarly, the model
is statistically misspecified at $P$ if there exists no value of $\psi$ that solves the
moment conditions
\[
0\not\in \left\{E_P[g(X_i,\psi)]:\psi\in\Psi\right\}.
\] Since econometrically correct specification imposes an additional restriction $(\vartheta
 (\psi)=\theta)$ relative to statistically correct specification, we see that the model
 may sometimes be econometrically, but not statistically, misspecified. In particular, when
 the model is just-identified with $\dim(g(X_i,\psi))=k=p=\dim(\psi),$ the moment
 conditions involve the same number of equations and unknowns and will often, though not always,  have a
 solution.  In over-identified models with $k>p,$ by contrast, we have more moment
 equations than unknowns.
\end{exsq}

\begin{exsq}[Linear Causal Model, continued]
    
As noted by e.g. \citet{breunig2018specification}, %
the random coefficient model implies a large collection of over-identifying restrictions.  For instance, the model implies that 
\[
\var_P(Y_i\mid W_i)= \var(\alpha_i)+2W_i\cov(\alpha_i,\beta_i)+W_i^2\var(\beta_i),
\] and thus that the conditional variance of $Y_i$ given $W_i$ must be quadratic in $W_i.$
 There are many ways for this (and other over-identifying restrictions) to fail, and thus
 for the model to be statistically misspecificed, $P\not\in\mathcal{P}$.\footnote{The support of $W_i$ matters for this conclusion.  If we instead consider binary $W_i\in\{0,1\}$ (and took e.g. $\theta=E[Y_i(1)-Y_i(0)],$ which is again equal to $E[\beta_i]$ under the model) then any $P$ with finite, nonzero $\var_P(W_iY_i)$ and $\var_P((1-W_i)Y_i)$ is consistent with $\mathcal{P}$.}

Despite this rich collection of over-identifying restrictions, there exist forms of econometric misspecification which do not imply statistical misspecification.  For instance, there are ways that random assignment may fail which are statistically undetectable but nevertheless distort our conclusions about $\theta=E[\frac{\partial}{\partial w}Y_i(W_i)]$.\footnote{For instance, it could be that 
$W_i=1\{F_{\beta|\alpha}(\beta_i|\alpha_i)>\frac{1}{2}\}\tilde{W}_i$ for a continuously distributed variable $\tilde{W}_i$ independent of $(\alpha_i,\beta_i)$ and 
$F_{\beta|\alpha}$ the conditional distribution function for $\beta_i$ given 
$\alpha_i$.  Thus $W_i$ is set to zero when $\beta_i$ exceeds its conditional median given
$\alpha_i$, and the mean of $\beta_i$ cannot be recovered from the observed data.} 
\end{exsq}

To formally account for the possibility of misspecification, suppose we are willing to
assume that the true $(\theta,P)$ pair is contained in a \emph{nesting model} $
\mathcal{M}^*$ which satisfies $\mathcal{M}\subset\mathcal{M}^*,$ and has a corresponding
 statistical model $\mathcal{P}^*.$  This nesting condition is vacuously satisfied when
 we consider the maximal nesting model $\mathcal{M}^*=\Theta\times\Delta(\mathcal{X})$ in which
 all distributions are consistent with all parameter values, but to obtain useful
 results we will need to impose further restrictions. The econometrics literature
 uses nesting models for two primary purposes:
\begin{enumerate}
\item To explore what statements can be made about $\theta$ under the nesting model $
\mathcal{M}^*.$ For instance, we could characterize and estimate the identified set for
 $\theta$ under $\mathcal{M}^*$, or derive confidence intervals which ensure coverage for
 $\theta$ under all $(\theta,P)\in\mathcal{M}^*$.  In many cases the strength of the
 conclusions obtained will depend on the details of the nesting model considered, so it
 is also common to explore sensitivity to the choice of nesting model $\mathcal{M}^*.$
\item To explore how specific estimators, motivated by the initial model $\mathcal
 {M}$, perform under the nesting model.  This could include interpreting the parameters
 these estimators recover for $(\theta,P)\in\mathcal{M}^*\setminus\mathcal{M}$ and
 deriving standard errors and confidence intervals which are valid for inference on these
 estimands.
\end{enumerate} 
This article focuses on the latter approach. 

\subsection{An aside on the many meanings of models}

Our definition of a statistical model follows a large literature in semiparametric statistics and econometrics, e.g. Chapter 25 of \cite{van2000asymptotic} and \cite{chen2018overidentification}.
By contrast, phrases like ``economic model'' and ``econometric model'' are used in many different
ways in the literature. %
Here we offer an incomplete survey and briefly relate some other definitions to ours. 
Our definition is in a sense minimal so long as one wishes to describe the properties of estimators for $\theta$. For instance, if we wish to characterize the set of parameter-value and bias pairs, $(\theta,\E_P[\hat\theta]-\theta),$ implied by any possible estimator $\hat\theta$ for $\theta,$ it is necessary and sufficient to know the set of possible values for the pair $(\theta,P).$ This, however, is precisely our definition of the econometric model.

A classical definition of a structural model is given by \citet{koopmans1950identification}. They
define a \emph{structure} $Q$ as a particular joint distribution over latent and observed
variables.\footnote{Precisely, \citet{koopmans1950identification} define a structure as a
distribution of latent variables and an equation linking observed and latent variables.} A
\emph{model} $\mathcal Q$ in \citet{koopmans1950identification} is then a set of
structures. The observed variables follow a distribution $P = P(Q)$, generated from a
structure $Q
\in
\mathcal Q$. A \emph{structural parameter} is a functional of the structure $\theta=
\theta(Q)$. In our notation, a structural model in the
\citet{koopmans1950identification} sense implies the econometric model \[
    \mathcal M = \br{(\theta(Q), P(Q)) : Q \in \mathcal Q }.
\]
To permit misspecification, we may consider a nesting model $\mathcal M^*$ generated by a
larger set of structures $\mathcal Q^* \supset \mathcal Q$.

Potential outcomes notation and structural causal models also provide natural nesting
models
\citep{splawa1990application,rubin1974estimating,pearl2009causality,richardson2013single,heckman2015causal}.
Many parameters of economic interest are functionals of the joint distribution $Q$ of all potential
outcomes and treatments, suitably defined.\footnote{In particular, structural equation models are often
generative models for potential outcomes. Relating to \citet{koopmans1950identification},
one could think of the joint distribution of potential outcomes and treatment variables as
the structure.} The distribution of potential outcomes then implies a distribution of the
observed data $P(Q)$. Our approach nests this as well: $
     \mathcal M = \br{(\theta(Q), P(Q)) : Q \in \mathcal Q}
 $ constitutes  a model in our sense.

These notions are clearer and more versatile than some others. Some textbooks
and reviews define an econometric model as deterministic economic relations plus
stochastic disturbances \citep{intriligator1983economic,woolridge2009introductory},
borrowing from the simultaneous equations literature. This usage is ambiguous, however, because the
error term's meaning is unclear: For instance, the equation
\[ 
Y_i = \alpha + \beta W_i + \epsilon_i
\]
may model linear, homogeneous causal effects $Y_i (w) = \alpha + \beta w + \epsilon_i$; or
it may simply define $\beta$ as a statistical parameter equal to the population OLS
coefficient, leaving the joint distribution of $(Y, W)$ unrestricted.\footnote{Foundational works in econometrics such as \cite{haavelmo1943simultenous, haavelmo1944econometrics} and \cite{koopmans1950identification} clearly distinguish these possibilities.  See e.g. \cite{pearl2015haavelmo} for discussion.} Under our
terminology, the two cases correspond to distinct econometric models.\footnote{The former
describes a model of homogeneous and linear potential outcomes \[
    \mathcal M = \br{(\beta, P) : \bigtimes_{i=1}^n(Y_i(W), W_i) \sim P, Y_i(w_1) - Y_i(w_2) = \beta (w_1 -
    w_2)}. 
\]
The latter defines a model in which \[
    \mathcal M = \br{\pr{\frac{\cov_P(W,Y)}{\var_P(W)}, P}: \bigtimes_{i=1}^{n}(Y_i, W_i)
    \sim P, \text{ $P$
    has second moments}}.
\]
}

\section{Choice of Estimator and Estimand}\label{sec: estimands}

Consider estimators $\hat\theta(X^n)\in\Theta$ that are consistent
under the model, i.e., for all $(\theta,P)\in\mathcal{M},$ $\hat\theta\pto \theta$ as
$n\to\infty$.  Under standard regularity conditions on the nesting model $\mathcal{M}^*$,
$\hat\theta$ still converges in probability to some limit when $P\in\mathcal{P}^*\setminus\mathcal{P}$. Denote this limit by $\plim(\hat\theta,P).$ Because
$\hat\theta$ is consistent under the model, $\plim(\hat\theta,P)=\theta^{*}(P)$ for every $P\in\mathcal
P$.  However, different choices of estimator will typically imply different limits $\plim
(\hat\theta,P)$ for $P$ outside the initial model, $P\in
\mathcal{P}^*\setminus \mathcal{P}$.\footnote{Indeed, as noted by \citet{chen2018overidentification} %
there is a close connection between testability of a statistical model $\mathcal{P}$ and the existence of distinct estimators for statistical functionals defined on the model.}  Thus, when choosing which estimator to use under the model $\mathcal{M},$ we are also (implicitly) choosing which parameter we wish to estimate when the model is statistically misspecified.

\begin{exsq}[Linear Causal Model, continued]
    
Under the model $\mathcal{M}$, we may estimate $\theta$ using the ordinary least squares
(OLS) estimator $\hat\theta_{\text{OLS}}=\frac{\cov_n(W_i,Y_i)}{\var_n(W_i)},$ for
$\cov_n$ and $\var_n$ the sample covariance and variance, respectively.  Alternatively, we
may instead estimate $\theta$ using the weighted least squares (WLS) estimator
$\hat\theta_{\text{WLS}}=\frac{\cov_{n,\omega}(W_i,Y_i)}{\var_{n,\omega}(W_i)},$ where $\cov_{n,\omega}$ and $\var_{n,\omega}$ weight observation $i$ by $\omega_i=\omega(W_i)$ for a weight
function $\omega.$ By the Gauss--Markov theorem, the efficient estimator uses $\omega_i=
\var_P(Y_i \mid W_i)^{ -1}$.  Consistency implies that $\plim(\hat\theta_
{\text{OLS}},P)=\plim(\hat\theta_{\text{WLS}},P)$ for all $P\in\mathcal{P}.$  By contrast,
under statistical misspecification $\plim(\hat\theta_{\text{OLS}},P)=
  \frac{\cov_P(W_i,Y_i)}{\var_P(W_i)}$ while $\plim(\hat\theta_{\text{WLS}},P)=\frac
   {\cov_{P,\omega}(W_i,Y_i)}{\var_{P,\omega}(W_i)},$ where
   these limits can differ for $P\not\in\mathcal{P}.$ Hence, once we allow for
   misspecification, WLS is not more efficient than OLS: The two simply estimate different
   things.
\end{exsq}

\begin{exsq}[Generalized Method of Moments, continued]
\label{ex:gmm_estimator} Moment condition models are usually estimated by the generalized
 method of moments (GMM). For $g_n(X^n,\psi)=\frac{1}{n}\sum g(X_i,\psi)$ the sample
 average moments and $\hat\Omega=\Omega_n(X^n)$ a data-dependent weighting matrix, GMM
 takes $\hat\theta_{\Omega}=\vartheta\left(\hat\psi_{\Omega}\right)$ for \[
\hat\psi_\Omega=\hat\psi_\Omega(X^n)=\argmin_{\psi\in\Psi}g_n(X^n,\psi)'\hat\Omega g_n(X^n,\psi).\]
For instance, the two-step efficient GMM estimator, which minimizes the large-sample
variance of the estimate under the model,  sets $\hat\Omega$ equal to the inverse of an estimate of the asymptotic variance of $\sqrt{n}g_n(X^n, \psi).$

If we define $\Omega_P=\plim (\hat\Omega,P)$---assumed to exist---and $g_P(\psi)
=
 \E_P[g(X_i,\psi)],$ then under mild conditions (see \citealt{imbens1997one,hall2003large}), %
 \[\plim(\hat\psi_\Omega,P)=\argmin_{\psi\in\Psi}g_P(\psi)'\Omega_Pg_P(\psi) \text
  { and }\plim (\hat\theta_\Omega,P)=\vartheta(\plim(\hat\psi_\Omega,P)).\] When the
  model is statistically correct, the argmin of the population GMM objective does not depend on the
  weighting matrix, and thus neither does the limit of $\hat\theta_\Omega$.  In contrast,
  when the model is statistically misspecified different weighting matrices yield
  different estimands. Hence, under misspecification the two-step GMM estimator is no
  more efficient than any other estimator: Once again each weighting leads us to recover
  a different parameter.
\end{exsq}

\begin{exsq}[Parametric models]
Previously, the estimators we discussed differed only by how they weighted a common set of objects ($W_i$
values in one case and moments in the other).  In other settings, estimators exploit the
data in qualitatively different ways. To illustrate, suppose we have a parametric model
which implies that the data are drawn from $P_\psi$, for a finite-dimensional
parameter $\psi,$ where $\theta=\vartheta(\psi)$. Assume $P_\psi$ is a continuous
distribution with density $p_\psi$, and that $X_i$ is identically distributed. Our econometric model is thus $\mathcal{M}=\left\{
(\vartheta (\psi),
P_\psi):\psi\in\Psi\right\},$ while our statistical model is
$\mathcal{P}=\left\{ P_\psi:\psi\in\Psi\right\}.$

Under the model $\mathcal{M}$ we may estimate $\theta$ by maximum likelihood, taking

\[
\hat\psi_{\text{MLE}}=\argmax_{\psi\in\Psi}\log p_\psi(X^n)
\]
and
$\hat\theta_{\text{MLE}}=\vartheta\left(\hat\psi_{\text{MLE}}\right).$ Alternatively, we may estimate
the model by the method of simulated moments, choosing a set of $\mathbb{R}^k$-valued functions
$h$, constructing moments comparing the realized values of
$h(X_i)$ to their model-implied means, $g(X_i,\psi)=h(X_i)-\E_{P_\psi}[h(X_i)],$ and
computing GMM estimates $\hat\psi_\Omega$ and $\hat\theta_\Omega$ for some
choice of weighting matrix. Provided the moment equations have a unique solution under
$P_\psi$ for all $\psi\in\Psi,$ these estimators have the same probability limit under the
model but maximum likelihood will be (weakly) more efficient.

As previously discussed, under misspecification and provided regularity conditions hold,
the GMM estimator $\hat\psi_\Omega$ converges to the value which minimizes the
$\Omega_P$-weighted deviation of the moment conditions from zero. In contrast,
$\plim(\hat\psi_{\text{MLE}},P)=\lim_{n\to\infty}\argmax_ {\psi\in\Psi} \E_P[\log p_\psi(X^n)],$ which
minimizes the Kullback--Leibler divergence between the distribution of the data and the
model-implied distribution \citep{white1982maximum}.
\end{exsq}

\smallskip

Once we allow that the model $\mathcal{M}$ is econometrically misspecified, we will usually
find that none of the estimands we consider recovers the precise parameter of economic
interest $\theta$.\footnote{For one thing, if $\theta$ is only set-identified under the
nesting model $\mathcal{M}^*$, then it is by definition impossible to construct an
estimator that consistently recovers $\theta$.}  Nevertheless, there are numerous results
in the literature which characterize the interpretation of specific estimands under
particular nesting models.  If one of these interpretations is ``close'' to the parameter
of interest $\theta$ in a qualitative sense, this may increase the appeal of the
corresponding estimator.

\begin{exsq}[Linear Causal Model, continued]
    The results of \citet{yitzhaki1996using} %
imply that the least-squares estimand $\plim(\hat\theta_{\mathrm{OLS}},P)$ can be written as a
weighted average of the partial derivative of $\E_P[Y_i \mid W_i]$ with respect to
$W_i,$
\[
\plim(\hat\theta_{\mathrm{OLS}},P)= \E_P\left[\eta_P(W_i)\frac{\partial}{\partial w} \E_P
[Y_i \mid W_i]\right]
\]
for particular weights $\eta_P(W_i) \ge 0$ with $\E_P[\eta_P(W_i)]=1.$  If the nesting
model $ \mathcal{M}^*$ maintains random assignment and differentiability of potential
outcomes $Y_i(w),$ it follows that $\plim(\hat\theta_{\mathrm{OLS}},P)=\E_P[\eta_P(W_i)
\frac{\partial}{\partial w}Y_i(W_i)],$ so OLS recovers a weighted average of causal
 effects, though typically not the unweighted average specified by $\theta.$  One can
 show an analogous result for weighted least squares, so the choice between the OLS and
 WLS estimands reduces to preference between the two sets of weights.
\end{exsq}

\smallskip

\begin{exsq}[Parametric Models, continued]
    In some economic applications of parametric models, researchers appear to prefer the
estimates from simulated method of moments to maximum likelihood.\footnote{For instance,
\citet{bordalo2020overreaction} write ``We prefer this method to maximum likelihood for
 two reasons. First, one advantage of our model is that it is simple and transparent.
 However, this simplicity comes at the cost of likely misspecification, and it is well
 known that with misspecification concerns moment estimators are often more reliable.''}
 This view has a long history in economics, with many researchers preferring SMM or
 related ``calibration'' procedures to maximum likelihood because of the ability to
 select moments that, in their judgment, more directly reflect the economic quantity of
 interest---see \citet{dawkins2001calibration} for an overview of calibration and
 extensive references.%

It is natural to interpret these statements as (implicit) restrictions on the nesting
model researchers have in mind, which ensure that some estimands are more closely related
to the parameters of interest than others or, relatedly, that some estimation criteria better reflect the 
researcher's objectives.  Such restrictions are rarely stated formally,
however.  Since these considerations appear to actively guide estimation choices, more
explicit articulation of the rationale for a given estimator is useful, and formalizing such
arguments seems a potentially fruitful area for future econometric research.
\end{exsq}

\begin{exsq}[Generalized Method of Moments, continued] 
As with OLS, the literature has characterized the GMM estimand for specific choices of
moments, weights, and nesting models.  For example, 
\citet{209996d6-9cfc-38cf-a7bb-28a4c5d811aa} and \citet{angrist1995two} show that, under
 monotonicity, two-stage least squares---a special case of GMM---recovers a
 positive-weight average of treatment effects.  %
 In finance, \citet{hansen1997assessing} %
show that in a misspecified asset pricing model, the GMM objective with a specific
weighting matrix measures the squared error in approximating an admissible stochastic
discount factor, and so measures the ``size'' of pricing errors in a precise sense.
\end{exsq}

As our discussion of parametric models suggests, even beyond settings covered by existing
theoretical guarantees, researchers' choice of weighting matrices for GMM and related
methods (e.g. simulated method of moments, minimum distance) are often informed
by concerns about misspecification.  To systematically explore this, we reviewed
empirical papers published in 
the \emph{American Economic Review} %
between 2020 and 2024 which, according to a keyword search, used moment-based methods.\footnote{We excluded methods like ordinary
least squares, two-stage least squares, and maximum likelihood which may be cast as
special cases of GMM but which the authors did not describe as such.} The search turned
up 36 papers.
We refer to these 
papers collectively as our ``GMM sample.''  In this sample, 22 of the papers were over-identified, and in 14 of these no results were reported based on an efficient weighting matrix.\footnote{In addition, some of the 
papers which notionally use efficient weights plug in parameters estimated in a first step.  This changes the form of the ``correct'' optimal weighting matrix, but is not accounted for in the implemented weighting matrices.  Consequently, the number of papers in fact using theoretically efficient weights is still lower.}  For those models where results based on efficient GMM are not reported, the most common (non-exclusive) choices were (i) the inverse of the diagonal of the asymptotic variance-covariance matrices (4 papers) (ii) identity weighting matrix (6 papers), and (iii) user-selected weighting matrices, often explicitly chosen based on domain knowledge (7 papers).

One reason the ``optimal'' weighting matrix may not be used in some cases is that in settings where the sample size is small relative to the number of moments, it is
 known that estimation of the optimal weighting matrix can lead to bias in both the
 estimate and the standard errors \citep[see e.g.][]{altonji1996small,windmeijer2005finite}.
One suggestion in such cases is to use the inverse of the diagonal alone, which 
\citet{altonji1996small} %
found yielded ``a substantial bias reduction'' in their problem.\footnote{Consistent with
 this interpretation, \citet{abebe2021selectionoftalent}
write that ``In line with the recent literature, we use this simple weighting matrix
 instead of the theoretically optimal weighting matrix, which may suffer from small sample
 bias \citep{altonji1996small}.''}  However, together with optimal weighting this still covers only about half of
 the over-identified papers in our GMM sample.  
 
 For the remaining two choices, identity
 weights and hand-selected weights, researchers are directly choosing the weight assigned
 to different moments, and thus the estimand.  This is most obvious for hand-selected
 weighting matrices, where in many cases the authors explicitly motivate their choice of weighting matrix based on their goals for estimation.\footnote{For instance, \citet{antras2020geography}
write that
``In order to guarantee that our model provides a proper quantitative evaluation of the general-equilibrium workings of the world economy, we place a higher weight on matching the empirical moments of larger trade flows.''
Similarly, \citet{einav2020screening}
write that ``Because identification is primarily driven by the discontinuous change in screening rates at age 40, we weight more heavily moments that are closer to age 40 than moments that are associated with younger and older ages.''}  Perhaps more subtly, for identity weighting matrices the scale that the authors choose for each moment (e.g. whether to measure a monetary outcome in dollars or thousands of dollars) determines the weight.\footnote{Weighting by the inverse of the diagonal of the variance eliminates dependence on the scale of individual moments, but is still sensitive to more general linear transformations of the moments, for instance replacing control and treatment means by control means and treatment-control contrasts.} In \cite{lise2020multi_skills}, for instance, the authors present a standard error formula based on identity weighting (see their Footnote 31), while their code uses a hand-selected diagonal but non-identity weighting matrix (which is consistent with the discussion in their footnote under an appropriate choice of units for the moments).

\begin{rmksq}[Further Results on Estimand Characterization] 

A growing literature characterizes estimands obtained from misspecified models in additional
settings.  An early positive result by \cite{angirst1998veterans} shows that in linear
models where a binary treatment is as good as randomly assigned given a discrete
covariate, regressions on the treatment and saturated dummies for the covariate recover a
positive weighted average of causal effects. More recently, however, a large literature
warns that estimands from linear (regression or instrumental variables) models fail to
recover positively weighted averages of causal effects in important cases. Examples
include difference in difference and event-study models estimated via two-way fixed effect
regressions \citep{chaisemartin2020twfe,callaway2020twfe,sun2021twfe},  IV models
estimated by two stage least squares regressions with multiple endogenous variables or
insufficiently rich controls
\citep{sloczynski2020should, bhuller2024tsls,blandhol2025tsls}, nonlinear estimators for
linear IV, such as LIML and continuously updating GMM
\citep{kolesar2013iv,andrews2019structure}, and linear regressions including both controls
and multiple treatments \citep{goldsmithpinkham2024contamination}.
\cite{chen2025potentialweights} develops a general method for characterizing the causal
estimand, if any, recovered by linear regression estimates.

Other recent papers characterize estimands in nonlinear models, including
\cite{Andrews2025structural} who study a class of nonlinear structural models and provide
necessary and sufficient conditions for instrumental variables estimates to recover linear
transformations of causal effects, as well as \cite{rambachan2025commontimeseriesestimands} and \cite{kolesar2025dynamiccausaleffectsnonlinear} who
consider estimation of causal effects in time-series models and derive positive results
for certain linear estimators, along with negative results for some popular nonlinear estimators.  Other recent work extends the concept of a pseudo-true value to set-identified models \citep{kwon2024misspecified, kaido2024informationbasedinferencemodels}.
\end{rmksq}

\section{Inference on Estimands}\label{sec: inference on estimands}

After fixing the estimator—and hence a statistical functional $\theta^*
(P)$ defined on $\mathcal{P}^*$---we naturally seek to quantify sampling uncertainty, but misspecification can
complicate inference. Under misspecification, standard error formulas derived under
$\mathcal P$ may no longer be correct; corresponding confidence intervals may undercover
the functional $\theta^*(P)$ when $P \in \mathcal P^* \setminus
\mathcal P$. This is a well-known motivation for misspecification-robust standard errors
for maximum likelihood \citep{white1982maximum}.

Similar results exist for GMM \citep{imbens1997one,hall2003large,hansen2021inference}, yet
they are seldom used in practice (analytic misspecification-robust standard errors do not appear to be used in any of the papers in our GMM sample).  Since these results seem to be less
well-known, we sketch an asymptotic expansion illustrating how misspecification impacts
GMM, as well as the intuition for robust standard errors. Let $\hat\psi_\Omega$ be the GMM
estimator with weighting matrix $\hat\Omega = \Omega + o_P (1)$ and $\psi^*_\Omega$ be its corresponding statistical functional.\footnote{Since we
consider behavior at a fixed $P,$ where possible we suppress the dependence on $P$ to
simplify notation.} Under statistically correct specification, $\psi_\Omega^*$ is simply the GMM
parameter defined by $\E_P [g (X_i,
\psi_\Omega^*)] = 0$ (assumed unique). Under misspecification, $\psi_\Omega^*$ minimizes the
population GMM criterion $\phi \mapsto (\E_P [g (X_i, \phi)])' \Omega \E_P [g (X_i,\phi)]$.
Let $\Gamma = \E_P[\nabla g(X_i, \psi_\Omega^*)]$ and
\[
\hat\Gamma = \frac{1}{n}
\sum_{i=1}^n \nabla g
(X_i, \hat\psi_\Omega) = \Gamma +O_P(1/\sqrt{n}).
\] By the first-order conditions for GMM, $\hat\Gamma' \hat\Omega
g_n(\hat\psi_\Omega) = 0 = \Gamma' \Omega \E_P[g_n(\psi_\Omega^*)]$.

To derive the standard errors for the GMM estimator under misspecification, note that by a first-order Taylor expansion and $\sqrt{n}$-consistency of $\hat\psi_\Omega,$
\begin{align*}
\sqrt{n} g_n(\hat\psi_\Omega) &=  \sqrt{n} g_n(\psi_\Omega^*) + 
    \underbrace{\pr{\frac{1} {n}\sum_
        {i=1}^n \nabla  g(X_i, \psi_\Omega^*)}}_{\Gamma + o_P(1)} \sqrt{n} (\hat\psi_\Omega - \psi_\Omega^*) + o_P
        (\sqrt{n}
    \norm{\hat\psi_\Omega-\psi_\Omega^*})  \\
    &= \sqrt{n} g_n(\psi_\Omega^*) + 
    \Gamma \sqrt{n} (\hat\psi_\Omega - \psi_\Omega^*) + o_P
        (1). \numberthis \label{eq:gmm_expansion}
\end{align*}
Rearranging and multiplying by $\hat\Gamma'\hat\Omega,$
\begin{equation}
        -(\hat\Gamma' \hat\Omega \Gamma) \sqrt{n}(\hat\psi_\Omega - \psi_\Omega^*) =  \sqrt{n} \Gamma' \Omega
 g_n(\psi_\Omega^*) + \sqrt{n} (\hat\Gamma' \hat\Omega - \Gamma' \Omega) g_n(\psi_\Omega^*) + o_P(1). \numberthis
 \label{eq:gmm_expansion_se}
 \end{equation}

When the model is well-specified $g_n(\psi_\Omega^*)$ has mean zero, so the second term in
\eqref{eq:gmm_expansion_se} is negligible, $ \sqrt{n} (\hat\Gamma' \hat\Omega - \Gamma'
\Omega) g_n(\psi_\Omega^*) = o_P(1)$, and the estimator is asymptoticaly normal with the conventional GMM asymptotic
variance:
\begin{align*}
\sqrt{n}(\hat\psi_\Omega - \psi_\Omega^*) &= -(\hat\Gamma' \hat\Omega \Gamma)^{-1}  \sqrt{n} \Gamma' \Omega g_n(\psi_\Omega^*) + o_P
    (1) \\
    &\dto \Norm(0, (\Gamma' \Omega \Gamma)^{-1} \Gamma'
    \Omega \Sigma(P) \Omega \Gamma (\Gamma' \Omega \Gamma)^{-1}),
\end{align*}
where $\Sigma(P)\equiv \avar_P(\sqrt{n}g_n(\psi_\Omega^*))$ for  $\avar_P(\cdot)$ the asymptotic variance under $P$.
However, when the model is statistically misspecified $\sqrt{n} (\hat\Gamma' \hat\Omega - \Gamma' \Omega) g_n(\psi_\Omega^*)$
is non-negligible and, under regularity conditions,
converges to a mean-zero Gaussian with nonzero, consistently estimable variance.  Thus, the familiar
sandwich formula is no longer correct and misspecification-robust standard errors should be used instead.

Since many economic models are statistically misspecified, it is questionable whether the conventional
GMM formula accurately captures the sampling variability of GMM estimators in practice.
Parallel to \citet{white1982maximum}'s widely-adopted suggestion for maximum likelihood, we recommend that robust standard errors be used for
GMM as well. Again parallel to maximum likelihood, robust standard errors for GMM do no harm, at least in large samples: They are equal
to the conventional standard errors when the model happens to be statistically well-specified, but remain valid for inference on $\psi_\Omega^*(P)$ when $P \not\in \mathcal {P}$.  Consequently, provided $\vartheta(\psi)$ is smooth in $\psi,$ delta-method standard errors will be valid for inference on $\theta^*_\Omega(P)=\vartheta(\psi_\Omega^*(P))$.

The robust standard error formula
\citep[e.g.][]{hall2003large} depends on additional terms not present in the usual formula, and can be complicated in some cases though software implementations are available.\footnote{For instance in the dcivreg command in Stata, which implements the misspecification-robust approach of \cite{Hwangetal2022} for linear IV.} An alternative for settings
with i.i.d. data is the standard nonparametric bootstrap, which approximates the limit
distribution in
\eqref{eq:gmm_expansion_se} when $\theta^*(P)$ varies smoothly with $P$.\footnote{As discussed by \cite{Hong_Li_2025}, the smoothness requirements for $\sqrt{n}$ asymptotic normality of GMM under misspecification are stronger than those for the correctly specified case.  The authors provide a bootstrap which is robust to insufficient smoothness.} For bootstrapping
GMM models, some propose recentering the bootstrapped moments so as to enforce the moment
restriction in the ``bootstrap world'' (\citealt{brown2002generalized};
also see Section 3.7 in \citealt{horowitz2001bootstrap}). When combined with a t-ratio
bootstrap, this approach generates asymptotic refinements under statistically correct specification.
Unfortunately, however, recentering yields invalid inference under statistical misspecification.
Intuitively, recentering imposes correct specification in the bootstrap data generating process.  Thus, the asymptotic variance of the estimator under the recentered bootstrap matches the conventional (invalid) sandwich formula.\footnote{Similarly, bootstrapping the moments but using the conventional formula to approximate the estimator variance, as done in some papers in our GMM sample, yields non-robust standard errors.}
We thus recommend \emph{against} recentering moments. 

\begin{rmksq}[A Bayesian Interpretation of Bootstrap Percentile Intervals]

While the asymptotic expansion \eqref{eq:gmm_expansion_se} is valid under misspecification, both the expansion itself and the resulting standard errors rely on regularity conditions, including the existence of a unique interior solution $\psi^*_\Omega(P)$ in the population GMM problem, strong identification of $\psi^*_\Omega(P)$, and differentiability of $\vartheta$.  Even when these conditions fail, however, the nonparametric bootstrap distribution for the estimate $\hat\theta_\Omega$ is asymptotically equivalent to the Bayesian posterior distribution for the estimand $\theta_\Omega^*(P)$ under a particular nonparametric and non-informative limit of priors.\footnote{See e.g. \cite{Chamberlain01012003, andrewsshapiro2024bootstrap} for discussion of the relevant priors, and Appendix D of the latter paper for sufficient conditions for asymptotic equivalence.    If we use the Bayes rather than nonparametric bootstrap, then the equivalence is exact.}  Thus, if we use bootstrap percentile intervals for inference on $\theta_\Omega^*,$ they offer correct large-sample coverage for well-behaved estimands $\theta^*_\Omega(P)$ regardless of model misspecification, while also providing a ``fallback'' interpretation as a Bayesian credible set for $\theta^*_\Omega(P)$ in less well-behaved cases.  The bootstrap also provides a natural tool for diagnosing the failure of the conventional, non-misspecification-robust GMM standard errors: if the bootstrap distribution  appears approximately normal and centered at $\hat\theta_\Omega,$ but differs from the normal suggested by the conventional standard error, this suggests the expansion \eqref{eq:gmm_expansion_se} may be reasonable but that the conventional, non-misspecification-robust asymptotic approximation is not.  See \cite{andrewsshapiro2024bootstrap} for formal results on this approach to evaluating asymptotic approximations.
\end{rmksq}

\begin{rmksq}[Further Results on Misspecification-Robust Inference] 

While \cite{white1980robust} is primarily framed around heteroskedasticity robustness,
these standard errors for linear regression are also robust to nonlinearity of the
conditional mean function and so represent a very widely used application of the
misspecification-robust approach. More recently, \cite{Lee2018tsls} highlights the
invalidity of conventional GMM standard errors in two stage least squares with
heterogeneous effects and multiple instruments, and specializes the formula of
\cite{hall2003large} to this case.  \cite{lee2014bootstrap} shows that bootstrapping a
misspecification-robust t-statistic yields asymptotic refinements for inference on the GMM
estimand, offering an alternative to recentering, and \cite{hansen2021inference} derive
misspecification-robust standard errors for the iterated-to-convergence GMM estimator.
Beyond standard errors, misspecification also invalidates many identification-robust
procedures: \cite{kleibergen2025CUGMM} derive an identification- and
misspecification-robust confidence interval for the continuously updating GMM estimand.
Finally, \cite{mueller2013sandwich} considers Bayesian decision making in misspecified
models and shows that for decision problems concerning maximum likelihood estimands, decision rules
based on an artificial posterior constructed using robust standard errors asymptotically
dominate conventional Bayesian inference.
\end{rmksq}

\section{Local Misspecification}\label{sec: local misspec}

Our study of statistical properties (e.g. probability limits and standard errors) has so
far focused on fixed, or ``global'' misspecification, since we have held $P$ fixed and considered behavior as the sample size grows.  As $n$ grows, however, any fixed
distribution $P\notin\mathcal P$ becomes detectable with probability one under mild conditions. In a GMM
setting, for instance, J-statistics
\[
J=\min_{\psi\in\Psi} n g_n(\psi)'\hat\Sigma^{-1}g_n(\psi)
\]
diverge to infinity under global misspecification, rather than to their limiting
$\chi^2_{k-p}$ distribution under statistically correct specification
\citep{hansen1982gmm}.

In many economic applications the degree of statistical misspecification appears
to be roughly on the same order as statistical uncertainty.  For instance, J-statistics,
while larger than expected under correct specification, are not always vastly so.
Estimates from different weighting matrices may be statistically distinguishable from one another, yet numerically and substantively fairly close.  Motivated by this observation, %
and following \cite{newey1985misspec}, a growing econometrics literature studies models of \emph{local}
misspecification, which assume that the sample of size $n$ is generated by a
data-generating process $P_n$ such that at some fixed value $\psi$ for the GMM parameters,
\[
    \E_{P_n}\bk{
        g(X_i, {\psi})
    } = \frac{\eta}{\sqrt{n}}.
\]
Because standard errors shrink at rate $\sqrt n$, the misspecification magnitude matches
their scale. 

As with other local asymptotic approximations, for instance in the literature studying
weak instruments \citep{staiger1997weakiv}, we do not imagine that the data distribution
would mechanically change were we able to gather more data.  Instead, local
misspecification is a modeling device to ensure that our asymptotic approximation reflects
a salient feature of many finite-sample problems, namely that the apparent degree of
misspecification is of the same order as sampling uncertainty. Smaller misspecification
would vanish compared to sampling noise; larger misspecification would swamp---and
invalidate, as discussed in the last section---standard errors. Hence local
misspecification is the appropriate regime for studying trade-offs between estimand choice
and estimator precision.

Local misspecification also allows important analytical simplifications. Specifically, in
locally misspecified settings many statistics behave like their correctly-specified
analogs, save for the addition of extra bias terms.  In particular,
\cite[][Lemma 4]{newey1985misspec}  shows that under local misspecification the GMM estimate
is asymptotically linear in the moments evaluated at $\psi,$
\begin{equation}\label{eq: linearized estimate}
\hat\psi_{\Omega}-\psi=-(\Gamma'\Omega\Gamma)^{-1}\Gamma'\Omega g_n(\psi)+o_p\left(\frac{1}{\sqrt{n}}\right)
\end{equation}
where $\Gamma$ and $\Sigma$ represent the limits of $\Gamma(P_n)$ and $\Sigma(P_n)$ respectively.
Thus, the asymptotic expansion under local misspecification is the same as that in the correctly specified case.  In particular, conventional GMM standard errors remain valid under local misspecification.

The local misspecification framework has been used in several recent papers.  For
instance, \citet{andrews2017measuring} suggest researchers report estimates of
$-(\Gamma'\Omega\Gamma)^{-1}\Gamma'\Omega$ to give intuition for the behavior of their estimates, since this describes how misspecification in
the moments (i.e. different values of $\eta$) feeds forward into bias in the estimates. 
\cite{armstrong2021sensitivity} impose the further restriction that
$\theta=\vartheta(\psi),$ so the GMM model is econometrically correctly specified in the case
where $\eta=0,$ and derive optimal confidence intervals for $\theta$ under restrictions on
the possible values for $\eta.$ \cite{bonhomme2022minimizing} consider models with
additional structure where some model components may be locally misspecified, and derive
asymptotically optimal estimators under bounds on the degree of misspecification.
Similarly, for a class of moment-condition models imposing parametric assumptions,
\cite{christensen2023robustness} derive bounds on quantities of interest under both local
and global misspecification.

Despite this recent activity using the local misspecification framework, there remains
scope for new results in this area.  To illustrate, in the remainder of
this section we use the local misspecification framework to develop---to the best of
our knowledge---a new result, characterizing the relationship between the J-statistic and
the class of GMM estimates obtained by varying the weighting matrix.

\subsection{J-Statistics, not J-Tests}

Relatively few economics papers report J-statistics. In our GMM sample, for instance, only
3 out of the 22 over-identified papers report J-statistics, either in the paper or the
online appendix.\footnote{One additional paper, \cite{mueller2021jobseekers}, reports the value of the two-step SMM objective function at its minimum. Because they use the optimal weighting matrix estimated by bootstrap, this objective value is equal to the  J-statistic absent simulation error.} One reason may be that few economists expect their models to hold
exactly. Under that view, a non-rejection signals insufficient power of the J-test, not
correct specification.  This suggests a focus on ``economic'' rather than statistical
measures of model fit.  Consistent with this, 19 of the 22 over-identified papers in our
GMM sample report some version of an informal identification check, often by plotting the
model-implied values of a set of moments against either (i) the data moments that were
used to estimate the model or (ii) a ``held out'' set of moments not used in estimation,
which may be interpreted as estimation moments assigned zero weight by $\hat\Omega$. Both
of these approaches convey information about the extent to which the model's
over-identifying restrictions appear to hold in the sample, but lack the formal
statistical interpretation of the J-statistic.

Such informal checks, while potentially useful, are an imperfect substitute for
J-statistics.  To highlight this point, in this section we show that J-statistics are
useful for much more than J-tests.  As discussed in the introduction, the fact that
researchers may choose different weighting matrices due to misspecification concern
introduces an additional researcher degree of freedom.  We show below that (under local
misspecification) J-statistics capture, in a precise sense, the range of estimates,
t-statistics, and confidence sets a researcher may obtain by varying their choice of
weight matrix.  This is valuable information for readers, and we encourage wider reporting
of J-statistics in applications of GMM.

To better understand the behavior of J-statistics in practice, we attempted to calculate J
statistics for the papers in our GMM sample which did not already report them.  In this
exercise we were limited to those papers with accessible data and code, that reported standard errors, and where we could implement our numerical optimization routines and were confident in their convergence.  Subject to these restrictions, we
were able to compute J-statistics for an additional three papers beyond those which
initially reported them.  Table \ref{tab:Jstats} collects the J-statistics for these
papers, along with the original three.  Strikingly, for the 6 papers in our sample where
we were able to obtain J-statistics, in only one case is the model rejected at the 5\%
level (although three papers have specifications rejected at the 10\% level).  Also
striking, while the number of papers is limited, there is not obvious selection in which
papers report J-statistics.  Indeed, the papers for which we computed J-statistics have, on
the whole, larger p-values (and thus, in that sense, less evidence of misspecification)
than the papers which originally reported J-statistics.

\begin{table}[h]
\centering
\caption{J-statistics for papers in the ``GMM sample''}
\label{tab:Jstats}
\begin{threeparttable}
\input{tabJstats}

\begin{tablenotes}
\item %
\footnotesize \textit{Notes:} This table reports the square root of the J-statistics and associated degrees of freedom and p-values for six papers in our ``GMM sample.'' See the text for inclusion criteria.
\end{tablenotes} 
\end{threeparttable}
\end{table}

To develop our formal result on J-statistics, we consider classes of weighting matrices with minimum and maximum eigenvalue bounded below and above by $\kappa^{-1}$ and $\kappa,$ respectively, for $\kappa\in[1,\infty)$ a positive constant:
\[
\mathcal{W}_{\kappa}=
\left\{\Omega\in\mathbb{R}^{k\times k}:\Omega=\Omega',\kappa^{-1}\le\lambda_{\min}(\Omega)
\le \lambda_{\max}(\Omega)\le\kappa\right\}.
\]
Provided the linear approximation \eqref{eq: linearized estimate} holds uniformly over such classes as $n\to\infty,$ the J-statistic characterizes, in a precise sense, the extent to which GMM results vary across weighting matrix choices.\footnote{We limit attention to $\Omega\in \mathcal{W}_\kappa$ because there is no hope that the linear approximation is uniformly valid over classes whose closure includes e.g. weight matrices with severely reduced rank.}

\begin{restatable}{assumption}{assuJstat}
\label{assu: J stat}
\emph{
The sample moment conditions evaluated at $\psi$ are asymptotically normal as $n\to\infty,$ 
$\sqrt{n}g_n(\psi)\dto N(\eta,\Sigma)$.  
Equation \eqref{eq: linearized estimate} holds uniformly, in the sense that
\[
\sup_{\Omega\in\mathcal{W}_{\kappa}}\sqrt{n}\left|\hat\psi_{\Omega}-\psi+(\Gamma'\Omega\Gamma)^{-1}\Gamma'\Omega g_n(\psi)\right|\pto 0
\]
as $n\to\infty$ for all fixed $\kappa\ge1.$  Moreover, we have estimators
$\hat\Gamma_{\Omega}$ and $\hat\Sigma_{\Omega}$ such that $\hat\Gamma_{\Omega}\pto \Gamma$
and $\hat\Sigma_\Omega\pto\Sigma$ uniformly over $\mathcal{W}_{\kappa}$, where both limits
are full rank and do not depend on $\Omega,$ and $\sqrt{n}
\norm{g_n(\hat\psi_{\Sigma^{-1}})-g_n(\psi)+\Gamma(\hat\psi_{\Sigma^
{-1}}-\psi)}\pto
0.$ }
\end{restatable}

Let
\[
    \hat\sigma_\Omega^2=\frac{1}{n} \hat{h}' \left(\hat\Gamma_{\Omega}'\Omega\hat\Gamma_{\Omega}\right)^
    {-1}\hat\Gamma_{\Omega}'\Omega\hat\Sigma_\Omega\Omega\hat\Gamma_{\Omega}\left
    (\hat\Gamma_{\Omega}'\Omega\hat\Gamma_{\Omega}\right)^{-1} \hat{h} \quad \hat{h} = 
    \diff{\vartheta
    (\hat\psi_\Omega)} {\psi}
\]
denote the delta-method variance estimator for $\hat\theta_\Omega$, and $\hat\Theta\left(\kappa,\tau\right)$ the set of GMM estimates that are attainable with (i) $\Omega\in \mathcal{W}_{\kappa}$ and (ii) standard error less than $\sqrt{1+\tau^2}$ times that of the efficient GMM estimator,
\[\hat\Theta\left(\kappa,\tau\right)=
\left\{\hat\theta_\Omega:\Omega\in\mathcal{W}_{\kappa},\hat\sigma_\Omega^2\le (1+\tau^2)\hat\sigma_{\Sigma^{-1}}^2\right\}.
\]
Finally, let $\hat\Theta(\tau)$ be the efficient GMM confidence interval with critical value $\tau\sqrt{J}$
\[
\hat\Theta(\tau)=\left[\hat\theta_{\Sigma^{-1}}\pm \tau\sqrt{J} \hat\sigma_{\Sigma^{-1}}\right].
\]
One can show that these sets are asymptotically equivalent for large $\kappa.$
\begin{restatable}{prop}{propjstatasymptotic}
\label{prop:J-stat-asymptotic}
Under \cref{assu: J stat}, and for
\[d_H(A,B)=\max\left\{\sup_{a\in A}\inf_{b\in B}|a-b|,\sup_{b\in B}\inf_{a\in A}|a-b|\right\}
\]the Hausdorff distance between the sets $A$ and $B,$ for any $\varepsilon>0,$
\[
\lim_{\kappa\to\infty}\limsup_{n\to\infty} \P\left( \sqrt{n}\cdot d_H\left(\hat\Theta\left(\kappa,\tau\right),\hat\Theta(\tau)\right)>\varepsilon\right)=0.
\]
\end{restatable}
\Cref{prop:J-stat-asymptotic} states that, asymptotically, GMM estimates with, e.g.,
 approximately $10\%$ higher asymptotic variance than the efficient estimator form an
 interval around $\hat\theta_{\Sigma^{-1}}$ with radius $\sqrt{10\%} \cdot \sqrt
 {J} \cdot \hat\sigma_{\Sigma^{-1}}$.  Thus, if we consider the set of estimates that a
 researcher can obtain at a given standard error ``cost'' relative to the efficient GMM
 estimate, this set is (asymptotically) characterized by the J-statistic.  Note that
 while we state this result using the ``oracle'' efficient estimator $\hat\theta_{\Sigma^
 {-1}}$ and standard error $\hat\sigma_{\Sigma^{-1}}$ for simplicity, under \cref{assu: J
 stat} it is equivalent to use the feasible variance estimate $\hat\Sigma_{\Omega}$ for
 any fixed, full-rank $\Omega.$

This result has immediate implications for the behavior of t-statistics and confidence
intervals.

\begin{restatable}{cor}{cortstatmainasymptotic}
\label{cor:tstatmain-asymptotic}
Under \cref{assu: J stat}, consider the t-statistic for testing $H_0:\theta=\theta_0$ based on the
GMM estimator with weighting matrix $\Omega$:
\[t_{\Omega}(\theta_0) = \frac{\hat\theta_\Omega-\theta_0}{\hat\sigma_{\Omega}}.\] Then for all $\varepsilon>0,$
\[
\lim_{\kappa\to\infty} \limsup_{n\to\infty}\P\left(\left|\inf_{\theta_0\in\mathbb{R}}\sup_{\Omega\in\mathcal{W}_{\kappa}}|t_{\Omega}(\theta_0)| - \sqrt{J}\right|>\varepsilon\right)=0.
\]
\end{restatable}

This result shows that, asymptotically, a researcher can find a weighting matrix that yields an absolute t-statistic of at least $\sqrt{J}$ for any possible null hypothesis.  Hence, for instance, when $\sqrt{J}>1.96$ a nefarious researcher who is unconstrained in their choice of weighting matrix can reject any null hypothesis at the 5\% level. Consequently, in order to limit the scope for ``weight-hacking'' to yield statistically significant results, it must be that either J-statistics are small or that other restrictions constrain the choice of weight matrix.\footnote{Among other results, \cite{Honoreetal2020} characterize the derivative of $\sigma_\Omega^2$ with respect to elements of $\Omega.$ Their results may thus be useful for building intuition about which $\Omega$ satisfy $\hat\sigma_\Omega^2\le (1+\tau^2)\hat\sigma_{\Sigma^{-1}}^2$.}  

Absent restrictions on the weighting matrix, the J-statistic threshold necessary for a result to be ``weight-hacking''-proof at the 5\% level, $\sqrt{J}\le1.96$, does not scale with the degree of over-identification, and so can be much more stringent than conventional J-test critical values.  In Table \ref{tab:Jstats}, for instance, four of ten specifications, and 3 of 6 papers, have $\sqrt{J}>1.96$, despite only one specification being rejected at the 5\% level by the J-test. 

Restated in terms of confidence intervals, we obtain the following result.
\begin{restatable}{cor}{corcsasymptotic}
\label{cor:cs-asymptotic}
    Under \cref{assu: J stat}, for a critical value $c$, consider confidence sets
    centered around the GMM estimator $\hat\theta_\Omega$,
    $\hat C_\Omega(c) = [\hat
    \theta_\Omega \pm c \hat\sigma_\Omega ]$. Then for all $\varepsilon>0,$
    \[
    \lim_{\kappa\to\infty}\limsup_{n\to\infty}\P\left(\left|\inf \br{c \ge 0: \bigcap_{\Omega\in\mathcal{W}_{\kappa}} C_\Omega(c) \neq \emptyset }-\sqrt{J}\right|>\varepsilon\right)=0.
    \]
    The intersection point is asymptotically equal to the efficient GMM estimator: 
    \[\lim_{\nu\to 0}\lim_{\kappa\to\infty}\limsup_{n\to\infty}\P\left(\left|\sqrt{n}\cdot d_H\left(\bigcap_{\Omega\in\mathcal{W}_{\kappa}} C_\Omega(\sqrt{J}+\nu),
    \left\{\hat\theta_{\Sigma^{-1}}\right\}\right)\right|>\varepsilon\right)=0.\]
\end{restatable}

Hence, to ensure at least one point of overlap between the confidence sets of researchers
considering different weighting matrices, researchers would need to use critical values no
smaller than $\sqrt{J}.$  This result bears an interesting resemblance to one by
\cite{andrews2024pseudotrue}.  There, working in a Bayesian framework and studying
linear-in-parameters moment condition models, the authors show that confidence intervals
with width $\sqrt{J}$ have correct average coverage under a class of priors motivated by
the GMM criterion.

Our asymptotic results in this section follow from exact, finite-sample results in a model, discussed in the
Appendix, where the moments are linear in parameters and have known variance.
\cite{armstrong2021sensitivity} show that this linear model corresponds to the limit
experiment for weakly identified GMM.

\begin{rmksq}[Further Results on Local Misspecification] 
Other recent work includes \cite{kitamura2013robustness}, who establish a
local-misspecification-robustness property for a generalized empirical likelihood
estimator, and \cite{andrews2020informativeness} who use a local misspecification
framework to examine the relationship between structural estimates and other, non-moment
statistics, with the goal of helping researchers clarify what ``drives'' their
estimates. \cite{fessler2019shrinkage} show how one can use the implications of locally
misspecified economic models to improve unbiased but potentially noisy estimates,
while \cite{armstrong2024adaptingmisspecification} consider the problem of optimally
combining biased and unbiased estimators under local misspecification.
\end{rmksq}

\bibliographystyle{aer}
\bibliography{main}

\newpage
\appendix

\section{Normal Limit Problem}\label{sec: Limit problem}

\citet{armstrong2021sensitivity} show that the problem of estimating $\theta$ under local misspecification reduces, asymptotically, to that of observing  
$
(Y, \Gamma, \Sigma, h)$ where we wish to estimate $\theta = h'\phi$ and \[Y
= -\Gamma \phi + \eta + \Sigma^{1/2} \epsilon \quad \epsilon \sim \Norm(0, I_k).  \label
{eq:limit_problem}\numberthis
\]
Quantities associated with GMM have natural analogues in the limit problem 
\eqref{eq:limit_problem}, which we collect in \cref{tab:analog}. These quantities are
linked in the sense that the asymptotic behavior of GMM quantities (left-side of 
\cref{tab:analog}) agrees with the behavior of the limit problem analogues under 
\eqref{eq:limit_problem}. In particular, GMM
estimators for $\psi$ with (symmetric, positive definite) weight matrix $\Omega$ correspond to estimators of $\phi$ in the
form \[
    \hat\phi_\Omega= -(\Gamma'\Omega \Gamma)^{-1} \Gamma' \Omega Y.
\]
The finite-sample distribution of $\hat\phi_\Omega$ in the model \eqref{eq:limit_problem}
\[
\hat\phi_\Omega \sim \Norm(\phi - (\Gamma'\Omega \Gamma)^{-1} \Gamma' \Omega\eta,  
(\Gamma'\Omega \Gamma)^{-1} \Gamma' \Omega \Sigma \Omega
\Gamma  
(\Gamma'\Omega \Gamma)^{-1})
\]
is exactly the GMM limit distribution under local misspecification. 
The corresponding estimators for $\theta$ is then $\hat\theta_\Omega = h'\hat\phi_\Omega$. Similarly,
the J-statistic has counterpart \begin{align*}
    J &= \min_u \, (Y + \Gamma u)\Sigma^{-1} (Y+\Gamma u) \\
    &= (Y + \Gamma
    \hat\phi_{\Sigma^{-1}})\Sigma^{-1} 
    (Y + \Gamma \hat\phi_{\Sigma^{-1}}).
\end{align*}
These statistics have even simpler representations in a canonical reparametrization of
\eqref{eq:limit_problem}. See
\cref{lemma:reparam}.

\begin{table}[tb]
    \caption{Quantities in GMM estimation and their analogues in the limit problem \eqref{eq:limit_problem}}
    \label{tab:analog}
    \centering

    \begin{tabular}{cc}
    \toprule
   GMM problem & Analogue in the limit problem \eqref{eq:limit_problem}\\
    \midrule
    $\psi$ & $\phi$ \\
    $\theta = \vartheta(\psi)$ & $\theta = h'\phi$ \\
      Jacobian of moments, $\E_P\bk{\diff{}{\psi} g(X_i, \psi)}$   &    $\Gamma$      \\ 
      Variance of moments, $\var_P(g(X_i, \psi))$ & $\Sigma$ \\ 
      Gradient of $\theta$ w.r.t. $\psi$, $\diff{\vartheta}{\psi}$ & $h$ \\
      GMM estimator of $\psi$ with weight matrix $\Omega_P$ & $-(\Gamma' \Omega\Gamma)^{-1} \Gamma' \Omega
      Y$ \\
      GMM estimator of $\theta$ with weight matrix $\Omega_P$ & $-h'(\Gamma' \Omega\Gamma)^{-1} \Gamma'
      \Omega
      Y$
       \\
       J-statistic $\min_{\psi \in \Psi} g_n(\psi)' \var_P(g(X_i, \psi))^{-1} g_n(\psi)$ &
       $\min_u (Y + \Gamma u) \Sigma^ {-1} (Y + \Gamma u)$ \\
    \bottomrule
    \end{tabular}
\end{table}

\begin{rmksq}[The problem \eqref{eq:limit_problem} as a limit version of GMM] 
    \label{rmk:limit_problem}
To clarify the limit problem \eqref{eq:limit_problem}, we briefly sketch the
connection to the asymptotics of GMM. Suppose we are given some (nonrandom) $\tilde \psi$
with \[ -\phi \equiv \sqrt{n}(\tilde \psi - \psi) = O(1),
\]
that is, we are given a parameter value in a $\sqrt{n}$ neighborhood of the parameter of interest.

We next define a local parameter $\phi$, based on the deviation from $\tilde\psi$. Any estimator $\hat\psi$ of
$\psi$ induces an estimator $\hat\phi$ of $\phi$ since \[
    \sqrt{n}(\hat\psi - \psi) = \underbrace{\sqrt{n}(\hat\psi - \tilde\psi)}_
    {\equiv \hat\phi} + \underbrace{\sqrt{n}
    (\tilde\psi -
        \psi)}_{-\phi} = \hat\phi - \phi.
\]
For $h \in \R^p$ the gradient of $\vartheta(\cdot)$, we also have by linearization \[
    \sqrt{n} (\vartheta(\hat\psi) - \vartheta(\psi)) = h'(\hat\phi - \phi) + o_P
    (1)
    \numberthis \label{eq:theta_expansion}
\]
for $\hat\psi$ that are $\sqrt{n}$-consistent.

Given $\tilde \psi$, we can summarize the statistical information in the GMM problem with
the value of the moments at $\tilde \psi$, $g_n(\tilde \psi)$. By a Taylor expansion
analogous to \eqref{eq:gmm_expansion}, \[
\sqrt{n}g_n(\tilde\psi) = \sqrt{n}g_n(\psi) + \Gamma \sqrt{n} (\tilde \psi - \psi) + o_P
(1) = \sqrt{n}g_n(\psi) - \Gamma\phi + o_P(1). 
\]

Finally, by Assumption \ref{assu: J stat}, $\sqrt{n} g_n(\psi) \dto \Norm(\eta, \Sigma)$ for
$\Sigma$ the limit of $\var_P[g(\cdot, \psi)]$. Thus, the observed quantity $Y \equiv 
\sqrt{n} g_n
(\tilde\psi)$ has asymptotic distribution \[
    Y \sim \Norm(-\Gamma \phi + \eta, \Sigma).
\]
The above display, combined with \eqref{eq:theta_expansion}, is equivalent to the problem
\eqref{eq:limit_problem}.\footnote{Without local misspecification, terms like
$(\hat\Gamma'\hat\Omega - \Gamma_P' \Omega_P)\eta$ are no longer asymptotically negligible, since
$\eta$ grows with $n$.}
\end{rmksq}

In the limit problem \eqref{eq:limit_problem}, we obtain exact, non-asymptotic analogs of \cref{prop:J-stat-asymptotic}, \cref{cor:tstatmain-asymptotic}, and \cref{cor:cs-asymptotic}.

\begin{restatable}{prop}{propjstat}
\label{prop:J-stat}
    In the limit problem \eqref{eq:limit_problem}, let $\hat\theta_{\Omega} = h'\hat\phi_{\Omega}$ and assume $\Gamma, \Sigma, h$ all have full
    rank. For $\tau \ge 0$ and $\sigma^2_\Omega=\var(\hat\theta_\Omega)$, the set of GMM estimates for $\theta$
    with variance at most $ (1+\tau^2) \sigma_{\Sigma^{-1}}^2$ is (almost surely) an interval centered at
    $\hat\theta_{\Sigma^{-1}}$ with radius $\sigma_{\Sigma^{-1}}\tau \sqrt{J}$: \[
        \br{h'\hat\phi_\Omega: \sigma^2_\Omega \le (1+\tau^2) \sigma_{\Sigma^{-1}}^2} = \bk{\hat\theta_{\Sigma^{-1}} \pm
        \sigma_{\Sigma^{-1}}\tau \sqrt{J}}.
    \]
\end{restatable}

\begin{restatable}{cor}{cortstatmain}
\label{cor:tstatmain}

In the setup of \cref{prop:J-stat}, consider the t-statistic for $H_0:\theta = \theta_0$ of the
GMM estimator with weighting matrix $\Omega$:
\[t_{\Omega}(\theta_0) = \frac{h'\hat\phi_\Omega-\theta_0}{\sqrt{\var(h'\hat{\phi}_\Omega)}}.\] Then  \[
\inf_{\theta_0}\sup_{\Omega} |t_{\Omega}(\theta_0)| = \sqrt{J}.
\]
\end{restatable}

\begin{restatable}{cor}{corcs}
\label{cor:cs}
    In the setup of \cref{prop:J-stat}, for a critical value $c$, consider confidence sets
    centered around the GMM estimator $\hat\theta_\Omega$,
    $\hat C_\Omega(c) = [\hat
    \theta_\Omega \pm c \sigma_\Omega ]$. Then $\sqrt{J}$ is the smallest critical value such that
    all
    confidence sets intersect: \[
        \sqrt{J} = \inf \br{c \ge 0: \bigcap_{\Omega} C_\Omega(c) \neq \emptyset }. \numberthis 
        \label{eq:conf_set_sqrtj}
    \]
    The intersection point is the efficient GMM estimator: $\bigcap_{\Omega} C_\Omega(\sqrt{J}) =
    \{\hat\theta_{\Sigma^{-1}}\}$.
\end{restatable}

To prove these results, it is helpful to work with a canonical, normalized form of the limit experiment:
\[
    Y^* = \colvecb{2}{\hat\phi_{\Sigma^{-1}}}{Z} = \colvecb{2}{\phi}{0} + \eta^* + {\Sigma^*}^{1/2}
    \epsilon^* \quad \epsilon^*
    \sim \Norm(0,I_k). \numberthis \label{eq:canonical}
\]
for some block-diagonal $\Sigma^* = \diag(\Sigma^*_\phi, \Sigma^*_Z)$. 

\begin{restatable}{prop}{propcanonical}
\label{lemma:reparam}
    Suppose $\Gamma$ is full-rank and $\Sigma$ is positive definite. Then 
    \eqref{eq:limit_problem} is equivalent to
    \eqref{eq:canonical} in the sense that there
    is some known and invertible matrix $Q$ whereby \[
        Y^* = QY \quad \eta^* = Q\eta \quad \Sigma^* = Q \Sigma Q'. 
    \]
    Moreover, recall  that $\hat\theta_\Omega = -h' (\Gamma' \Omega \Gamma)^{-1} \Gamma' \Omega Y$,
    $\sigma_\Omega^2 = \var(\hat\theta_\Omega)$. Then, 
\begin{enumerate}
    \item GMM estimators can be written as $\hat\theta_\Omega = h'\hat\phi_\Omega= h'\hat\phi_{\Sigma^{-1}} +
    v_\Omega'Z$, where any $v_\Omega$ corresponds to some choice of the weighting matrix \[\br{v_\Omega :
    \Omega \text{ is positive definite}} = \R^ {k-p}.\]

    \item $\sigma^2_\Omega = \sigma_{\Sigma^{-1}}^2 + v_\Omega' \Sigma_Z^* v_\Omega$
    \item  The J-statistic is the squared norm of $Z$: $J = Z'\pr{\Sigma_Z^*}^{-1} Z$. 
\end{enumerate}
\end{restatable}

\subsection{Proofs for Normal Limit Problem}\label{sec: Limit problem proofs}

\propjstat*

\begin{proof}
Consider the reparametrization of \eqref{eq:limit_problem} in \eqref{eq:canonical}. \Cref{lemma:reparam} verifies that \begin{enumerate}
    \item GMM estimators can be written as $\hat\theta_\Omega = h'\hat\phi_\Omega= h'\hat\phi_{\Sigma^{-1}} +
    v_\Omega'Z$, where any $v_\Omega$ corresponds to some choice of the weighting matrix \[\br{v_\Omega :
    \Omega \text{ is positive definite}} = \R^ {k-p}.\]

    \item $\sigma^2_\Omega = \sigma_{\Sigma^{-1}}^2 + v_\Omega'\Sigma_Z^* v_\Omega$
    \item  The J-statistic is the norm of $Z$: $J = Z'{\Sigma_Z^*}^{-1} Z$. 
\end{enumerate}

As a result, for $v_{\Omega(Z)} = c {\Sigma^*_Z}^{-1} Z$, we have that $\hat\theta_{\Omega(Z)} =
\hat\theta_{\Sigma^{-1}} + c J$ and \[
    \sigma_{\Omega(Z)}^2 = \sigma_{\Sigma^{-1}}^2 + c^2 J.
\]
As we vary $c^2 \le \frac{\tau^2 \sigma_{\Sigma^{-1}}^2}{J}$, we obtain the range of estimates $
\bk{\hat\theta_{\Sigma^{-1}}
\pm
        \sigma_{\Sigma^{-1}} \tau \sqrt{J}}$. This proves \[
            \br{h'\hat\phi_\Omega: \sigma^2_\Omega \le (1+\tau^2) \sigma_{\Sigma^{-1}}^2} \subset 
            \bk{\hat\theta_{\Sigma^{-1}} \pm
        \sigma_{\Sigma^{-1}} \tau \sqrt{J}}. 
        \]
        On the other hand, note that \[
            \sup_{v_\Omega' \Sigma_Z^*   v_\Omega \le \tau^2 \sigma_{\Sigma^{-1}}^2}
            v_\Omega'Z  = \tau \sigma_{\Sigma^{-1}} 
            \norm{{\Sigma_Z^*}^{-1/2} Z} = \tau\sigma_{\Sigma^{-1}} \sqrt{J}
        \]
        Thus, anything outside of $\bk{\hat\theta_{\Sigma^{-1}} \pm
        \sigma_{\Sigma^{-1}} \tau \sqrt{J}}$ is not attainable by any choice of $v_\Omega$. This completes the
        proof. 
\end{proof}

\corcs*

\begin{proof}
    Suppose $c < \sqrt{J}$ and let $\tau > 0$. By the proof of 
    \cref{prop:J-stat}, there is some estimator
    $\hat\theta_\Omega$ where \[
        \hat\theta_\Omega - \hat\theta_{\Sigma^{-1}} = \sigma_{\Sigma^{-1}} \tau \sqrt{J} \quad \sigma_\Omega =
        \sqrt{1+\tau^2} \sigma_{\Sigma^{-1}}.
    \]
    The confidence set $C_\Omega(c)$ has radius $c \sqrt{1+ \tau^2} \sigma_{\Sigma^{-1}} < \tau
    \sqrt{J} \sigma_{\Sigma^{-1}}$ for sufficiently large $\tau^2$. Thus, for sufficiently large
    $\tau^2$, the confidence set $C_\Omega(c)$ would fail to include $\hat\theta_
    {\Sigma^{-1}}$. On the
    other hand, for such a $\tau$, there is another estimator that is on the other side
    of $\hat\theta_{\Sigma^{-1}}$: $\hat\theta_
    {\Omega'} = \hat\theta_{\Sigma^{-1}} - \sigma_{\Sigma^{-1}} \tau \sqrt{J}$. Hence $C_\Omega(c) \cap C_{\Omega'}(c) =
    \emptyset$ since neither extends to $\hat\theta_{\Sigma^{-1}}$. For $c = \sqrt{J}$, \[
        c \sqrt{1+\tau^2} \sigma_{\Sigma^{-1}} \ge \tau \sqrt{J} \sigma_{\Sigma^{-1}}
    \]
    for all $\tau \ge 0$.  Hence all confidence sets include $\hat\theta_{\Sigma^{-1}}$. This proves
    \eqref{eq:conf_set_sqrtj}.

    Given
    $\hat\theta \neq \hat\theta_{\Sigma^{-1}}$, without loss, suppose $\hat\theta < \hat\theta_{\Sigma^{-1}}$.
    Then for $\hat\theta_\Omega = \hat\theta_{\Sigma^{-1}} + \sigma_{\Sigma^{-1}} \tau \sqrt{J}$, \[
        |\hat\theta_\Omega - \hat\theta| = \sigma_{\Sigma^{-1}} \tau \sqrt{J} + \hat\theta_{\Sigma^{-1}} - \hat\theta 
        > \sqrt{J} \sqrt{1+\tau^2} \sigma_{\Sigma^{-1}}
    \]
    for sufficiently large $\tau$, since \[\sqrt{1+\tau^2} - \tau = O(\tau^{-1})\] as $\tau \to
    \infty$.  Thus the confidence set around $\hat\theta_\Omega$ would exclude $\hat\theta$.
    This proves that $\bigcap_{\Omega} C_\Omega(\sqrt{J}) = \{\hat\theta_{\Sigma^{-1}}\}$.
\end{proof}

\cortstatmain*

\begin{proof}
This is immediate from \cref{cor:cs}.  In particular, note that by \cref{cor:cs},
$|t_\Omega(\hat\theta_{\Sigma^{-1}})|\le\sqrt {J}$ for all $\Omega,$ while for any $\theta $
there exists $\Omega$ with $t_\Omega(\theta)\ge \sqrt{J}$.
\end{proof}

\propcanonical*

\begin{proof}

    Let $\Lambda = -(\Gamma' \Sigma^{-1} \Gamma)^{-1} \Gamma' \Sigma^{-1} \in \R^{p
    \times k}$ and let $M = \tilde M (I + \Gamma \Lambda)$ where $\tilde M$ is any matrix
    so that $M$ is full-rank. $\tilde M$ exists since $-\Gamma \Lambda$ is idempotent and
    of rank $p$. Thus $I + \Gamma \Lambda$ is of rank $k-p$. Let \[ Q =
    \colvecb{2}{\Lambda}{M}.
    \]

    By multiplication, it is easy to verify that \[
        Q^{-1} = [-\Gamma, \Sigma M'(M \Sigma M')^{-1}]. 
    \]
    It is also easy to verify that $M \Gamma = 0 = \Lambda \Sigma M'$. With this choice of
    $Q$, \[
        \Sigma^* = Q \Sigma Q' = \colvecb{2}{\Lambda \Sigma }{M \Sigma} [\Lambda', M'] = 
        \begin{bmatrix}
            \Lambda \Sigma \Lambda' & \\ 
            & M \Sigma M'
        \end{bmatrix}
    \]
 is diagonal, as desired. Note that \[
     QY = \colvecb{2}{\hat\phi_{\Sigma^{-1}}}{MY} \quad M \Gamma \phi = 0
 \]
 is of the desired form as well.

For the first claim after ``moreover,'' write \begin{align*}
    -(\Gamma' \Omega \Gamma)^{-1} \Gamma' \Omega Y &= - (\Gamma' \Omega \Gamma)^{-1}
    \Gamma' \Omega \colvecb{2}
    {\Lambda}{M}^{-1} \colvecb{2}{\hat\phi_{\Sigma^{-1}}}{Z} \\ 
    &=\hat\phi_{\Sigma^{-1}} - (\Gamma' \Omega \Gamma)^{-1} \Gamma' \Omega \Sigma M'( M
    \Sigma M')^{-1} Z.
    \end{align*}
    Thus, \[
        v_\Omega = -( M \Sigma M')^{-1} M \Sigma \Omega \Gamma (\Gamma' \Omega \Gamma)^
        {-1} h.
    \]
    
    To complete the proof of part (1) of the Proposition, the following is checked by
    \cref{lemma:oblique}: \begin{align*}
    \br{\Omega \Gamma (\Gamma' \Omega \Gamma)^{-1} h: \Omega} &= \br{q \in \R^k : \Gamma' q = h} 
        \label{eq:oblique_projection} \numberthis\\
        &=
        \br{q \in \R^k : q = q_0 + v, \Gamma'q_0 = h, v \in N(\Gamma')}.
    \end{align*}  From this, it follows that the set of $v_\Omega$, as $\Omega$ ranges over all weighting matrices, is \[
        \br{-( M \Sigma M')^{-1} M \Sigma q_0 - ( M \Sigma M')^{-1} M \Sigma v : v \in N
        (\Gamma')}.
    \]
    Thus, for (1), it suffices to show that $\br{M \Sigma v : v \in N(\Gamma')} = \R^
    {k-p}$. It further suffices to show that \[
        \br{\Sigma^{1/2} v : v \in N(\Gamma')} = N(M \Sigma^{1/2})^{\perp}.
    \]
    To that end, note that \begin{align*}
     \br{\Sigma^{1/2} v : v \in N(\Gamma')} &= \br{u : \Gamma' \Sigma^{-1/2} u = 0} = N
     (\Gamma' \Sigma^{-1/2})\\
     N(M \Sigma^{1/2}) &= \br{v : M \Sigma^{1/2} v = 0} = \br{v : \Sigma^{1/2} v \in R
     (\Gamma)} \tag{$N(M) = R(\Gamma)$}\\
     &=\br{\Sigma^{-1/2}\Gamma u : u \in \R^p} \\ 
     &= R(\Sigma^{-1/2} \Gamma) = N(\Gamma' \Sigma^{-1/2})^\perp.
    \end{align*}
    This verifies (1). 

    (2) follows immediately from the observation that $\Sigma^*$ is block-diagonal. 

    For (3), we may compute that \begin{align*}
    J &= \norm{
        Y + \Gamma \hat\phi_{\Sigma^{-1}}
    }_{\Sigma^{-1}}^2 \\
    &= \norm{Q^{-1}(QY + Q \Gamma \hat\phi_{\Sigma^{-1}})}_{\Sigma^{-1}}^2 \\ 
    &= \norm{QY + Q \Gamma \hat\phi_{\Sigma^{-1}}}_{(Q'\Sigma Q)^{-1}}^2 \\
    QY + Q \Gamma \hat \phi_{\Sigma^{-1}} &= \colvecb{2}{\hat\phi_{\Sigma^{-1}}}{Z} - \colvecb{2}
    {\hat\phi_{\Sigma^{-1}}}{0}
    =
    \colvecb{2}{0}{Z}\\
    Q'\Sigma Q &= \Sigma^*.
    \end{align*} 
    Finally, observe that \[
        \colvecb{2}{0}{Z}' {\Sigma^*}^{-1} \colvecb{2}{0}{Z} = Z' {\Sigma^*_Z}^{-1}
        Z.
    \]
    This completes the proof.
\end{proof}

\begin{lemma}
\label{lemma:oblique}
    In the proof of \cref{lemma:reparam}, \eqref{eq:oblique_projection} holds. 
\end{lemma}

\begin{proof}
    The $\subset$ direction in \eqref{eq:oblique_projection} is immediate.
     For the $\supset$ direction, fix some $q$ where $\Gamma' q = h$. 
     We can uniquely
     write \[ q = u + w
    \]
    where $u \in R(\Gamma)$  and $w \in N(\Gamma')$. Note that we can write $u$ as its
    own projection onto $R(\Gamma)$, $u = \Gamma
    (\Gamma'\Gamma)^{-1}
    \Gamma' u $. Since $\Gamma'q = \Gamma'u = h$, we have $u = \Gamma
    (\Gamma'\Gamma)^{-1}h$. We can decompose \[
        \R^k = R(\Gamma) \oplus {\br{cw : c\in \R}} \oplus V
    \]
    for subspace $V$ the orthocomplement to $R(\Gamma) \oplus \br{cw : c\in \R}$. For some
    $a > 0$, take \[ \Omega = \Gamma \Gamma' + a(uw' + wu') + aww' + P_V
    \]
    for $P_V$ the projection matrix onto $V$. Note that by construction, any nonzero
    vector $t$
    can be written as \[
        t = \Gamma t_1 + c w + t_2 \quad t_1 \in \R^p \quad t_2 \in V
    \]
   Set $a = \frac{1}{2h'(\Gamma'\Gamma)^{-2} h} > 0$, and  
    observe that $\Omega$ is positive definite: \begin{align*}
    t' \Omega t &= t_1 (\Gamma'\Gamma)^2 t_1 + a c^2 \norm{w}^4 + \norm{t_2}^2 + 2a c
        \norm{w}^2 t_1'h \\
        & > a (t_1'h)^2 + ac^2 \norm{w}^4 + 2ac \norm{w}^2 t_1'h + \norm{t_2}^2 
        \tag{By Cauchy--Schwarz, $(t_1'h)^2 \le (t_1' (\Gamma'\Gamma)^2 t_1)(h' (\Gamma'
        \Gamma)^ {-2} h)$
        }
        \\
        &= a
        (t_1'h + c\norm{w}^2)^2 + \norm{t_2}^2 \ge 0.
    \end{align*}
    That the first inequality is strict when $t'_1h\neq0$ is straightforward. If $t'_1h=0$, either $t_1\neq0$ and Cauchy-Schwarz is strict or $t_1=0$, implying that $c$ or $t_2$ are not zero. In any case, the strict inequality in display is justified. 

    For the choice $a=\frac{1}{2h'(\Gamma'\Gamma)^{-2}h,}$ observe that \[
        \Gamma' \Omega \Gamma = (\Gamma'\Gamma)^2 \qquad \Omega \Gamma = \Gamma 
        (\Gamma'\Gamma) + aw
        u'\Gamma = \Gamma (\Gamma'\Gamma) + a w h'
    \]
    and thus \[
\Omega \Gamma (\Gamma' \Omega \Gamma)^{-1} = \Gamma (\Gamma' \Gamma)^{-1} + a w h'
        (\Gamma' \Gamma)^{-2}
   \implies
    \Omega \Gamma (\Gamma' \Omega \Gamma)^{-1} h = \Gamma (\Gamma'\Gamma)^{-1} h + w = u+w = q
    \]
    This proves \eqref{eq:oblique_projection}. 
\end{proof}

\section{Proofs for Results in Main Text}\label{sec: proofs}

\propjstatasymptotic*
\begin{proof}

We have assumed that for all fixed $\kappa\ge1,$ and $\phi_{\Omega}^{*}=-\sqrt{n}\cdot(\Gamma'\Omega\Gamma)^{-1}\Gamma'\Omega g_{n}(\psi)$
\[
\sqrt{n}(\hat{\psi}_{\Omega}-\psi)-\phi_{\Omega}^{*}\pto 0,~~\hat{\Gamma}_{\Omega}\pto\Gamma,~~\hat{\Sigma}_{\Omega}\pto\Sigma
\] 
uniformly over $\Omega\in\mathcal{W}_{\kappa}.$ Since we have also
assumed $\vartheta$ is continuously differentiable and $\Gamma,\Sigma$
are full rank, it follows immediately that for $\hat{\theta}^{*}_\Omega=h'\phi^*_
{\Omega}$
\begin{equation}
\sqrt{n}(\hat{\theta}_{\Omega}-\vartheta(\psi))-\hat{\theta}^{*}_\Omega\pto 0,~~\sqrt{n}
\hat{\sigma}_{\Omega}\pto\sigma_{\Omega}=\pr{h'\left(\Gamma'\Omega\Gamma\right)^
{-1}\Gamma'\Omega\Sigma\Omega\Gamma\left(\Gamma'\Omega\Gamma\right)^{-1}h}^{1/2},
\label{eq:estimator
and se convergence}
\end{equation}
again uniformly over $\mathcal{W}_{\kappa}.$ Since $\Sigma^{-1}\in\mathcal{W}_{\kappa}$
for $\kappa$ sufficiently large (which we assume for the remainder of the proof), it
further follows that 
\[\sqrt{n}g_{n}\left(\hat{\psi}_{\Sigma^{-1}}\right)-\sqrt{n}g_{n}\left(\psi\right)-\Gamma\phi_{\Sigma^{-1}}^{*}\pto0,
\]
and that for
\[ 
J = n\cdot
g_{n}\left(\hat{\psi}_{\Sigma^{-1}}\right)'\hat{\Sigma}^{-1}g_{n}\left(\hat{\psi}_{\Sigma^{-1}}\right)
\]
and
\[
J^{*}=\min_{u}(\sqrt{n}g_{n}(\psi)+\Gamma u)'\Sigma^{-1}(\sqrt{n}g_{n}(\psi)+\Gamma u),
\]
where the right hand side is minimized at $\phi_{\Sigma^{-1}}^{*},$
$J-J^*\pto 0.$

Let us define analogs of $\hat{\Theta}\left(\kappa,\tau\right)$ and
$\hat{\Theta}(\tau)$ based on $\hat\phi_{\Omega}^{*}$, 
\[
\hat{\Theta}^{*}(\kappa,\tau)=\left\{ h'\hat{\phi}_{\Omega}^{*}:\Omega\in\mathcal{W}_{\kappa},\sigma_{\Omega}^{2}\le(1+\tau^{2})\sigma_{\Sigma^{-1}}^{2}\right\} 
\]
\[
\hat{\Theta}^{*}(\tau)=\left\{ h'\hat{\phi}^*_{\Sigma^{-1}}\pm\tau\sqrt{J^{*}}\sigma_{\Sigma^{-1}}\right\} .
\]
Note that by the triangle inequality and the definition of Hausdorff
distance,
\begin{align*}
&\sqrt{n}\cdot d_{H}\left(\hat{\Theta}\left(\kappa,\tau\right),\hat{\Theta}(\tau)\right)=d_
{H}\left(\sqrt{n}\cdot\left(\hat{\Theta}\left(\kappa,\tau\right)-\vartheta(\psi)\right),\sqrt{n}\cdot\left(\hat{\Theta}(\tau)-\vartheta(\psi)\right)\right)\\
&\le
d_{H}\left(\sqrt{n}\cdot\left(\hat{\Theta}\left(\kappa,\tau\right)-\vartheta(\psi)\right),\hat{\Theta}^{*}\left(\kappa,\tau\right)\right)+d_{H}\left(\hat{\Theta}^{*}\left(\kappa,\tau\right),\hat{\Theta}^{*}\left(\tau\right)\right)
\\
&\quad\quad+
d_{H}\left(\sqrt{n}\cdot\left(\hat{\Theta}(\tau)-\vartheta(\psi)\right),\hat{\Theta}^{*}(\tau)\right).
\end{align*}

We will show that the first and third terms on the right hand side
converge in probability to zero as $n\to\infty$ for fixed $\kappa$,
while the second term converges in probability to zero as $n\to\infty$ and $\kappa\to\infty$, in that order.

Let us begin with the third term. Since 
\[
\sqrt{n}\cdot\left(\hat{\Theta}(\tau)-\vartheta(\psi)\right)=\left[\sqrt{n}(\hat{\theta}_{\Sigma^{-1}}-\vartheta(\psi))\pm\tau\sqrt{J}\sqrt{n}\hat{\sigma}_{\Sigma^{-1}}\right],
\]
(\ref{eq:estimator and se convergence})
immediately implies that $\plim\left(d_{H}\left(\sqrt{n}\cdot\left(\hat{\Theta}(\tau)-\vartheta(\psi)\right),\hat{\Theta}^{*}(\tau)\right)\right)=0$.

Next, let us turn to the first term, and show that
\begin{equation}
\plim\left(d_{H}\left(\sqrt{n}\cdot\left(\hat{\Theta}\left(\kappa,\tau\right)-\vartheta(\psi)\right),\hat{\Theta}^{*}\left(\kappa,\tau\right)\right)\right)=0.\label{eq: first term convergence}
\end{equation}
To this end, define 
\[
\mathcal{\hat{W}}_{\kappa,\tau}=\left\{ \Omega\in\mathcal{W}_{\kappa}:\hat{\sigma}_{\Omega}\le\sqrt{\left(1+\tau^{2}\right)}\hat{\sigma}_{\Sigma^{-1}}\right\} 
\]
 and note that since 
\[
\sqrt{n}\cdot\left(\hat{\Theta}\left(\kappa,\tau\right)-\vartheta(\psi)\right)=\left\{
\sqrt{n}\left(\vartheta(\hat{\psi}_{\Omega})-\vartheta\left(\psi\right)\right):\Omega\in\mathcal{\hat{W}}_{\kappa,\tau}\right\}
,
\]
(\ref{eq:estimator and se convergence})
implies that
\[
d_{H}\left(\sqrt{n}\cdot\left(\hat{\Theta}\left(\kappa,\tau\right)-\vartheta(\psi)\right),\left\{ h'\hat{\phi}_{\Omega}^{*}:\Omega\in\mathcal{\hat{W}}_{\kappa,\tau}\right\} \right)\pto0.
\]
Hence, to prove the convergence (\ref{eq: first term convergence})
it suffices to show that 
\[
d_{H}\left(\left\{ h'\hat{\phi}_{\Omega}^{*}:\Omega\in\mathcal{\hat{W}}_{\kappa,\tau}\right\} ,\hat{\Theta}^{*}\left(\kappa,\tau\right)\right)\pto0.
\]
Note, however, that if we define 
\[
\mathcal{W}_{\kappa,\tau}=\left\{ \Omega\in\mathcal{W}_{\kappa}:\sigma_{\Omega}\le\sqrt{\left(1+\tau^{2}\right)}\sigma_{\Sigma^{-1}}\right\} ,
\]
then $\lim_{\delta\to0}d_{H}\left(\mathcal{W}_{\kappa,\tau},\mathcal{W}_{\kappa,\tau+\delta}\right)=0$
(where we now consider the Hausdorff distance defined based on e.g. Frobenius norm). Consequently, since (\ref{eq:estimator and se convergence})
implies that for $\delta>0$
\[
\P\left\{ \mathcal{W}_{\kappa,\tau-\delta}\subseteq\mathcal{\hat{W}}_{\kappa,\tau}\subseteq\mathcal{W}_{\kappa,\tau+\delta}\right\} \to1,
\]
we see that $d_{H}\left(\mathcal{W}_{\kappa,\tau},\mathcal{\hat{W}}_{\kappa,\tau}\right)\pto0.$
Note next that $\frac{\phi_{\Omega}^{*}}{\left\Vert \sqrt{n}g_{n}\left(\psi\right)\right\Vert }$
is a continuous function of $\left(\Omega,\frac{\sqrt{n}g_{n}\left(\psi\right)}{\left\Vert \sqrt{n}g_{n}\left(\psi\right)\right\Vert }\right)\in\mathcal{W}_{\kappa}\times\mathbb{S}^{k}$
for $\mathbb{S}^{k}$ the unit sphere in $\mathbb{R}^{k}$. Since
$\mathcal{W}_{\kappa}\times\mathbb{S}^{k}$ is compact, it follows
that $\frac{\phi_{\Omega}^{*}}{\left\Vert \sqrt{n}g_{n}\left(\psi\right)\right\Vert }$
is uniformly continuous. Convergence of $\mathcal{\hat{W}}_{\kappa,\tau}$
to $\mathcal{W}_{\kappa,\tau}$ thus implies that 
\[
\left\Vert \sqrt{n}g_{n}\left(\psi\right)\right\Vert ^{-1}d_{H}\left(\left\{ h'\hat{\phi}_{\Omega}^{*}:\Omega\in\mathcal{\hat{W}}_{\kappa,\tau}\right\} ,\hat{\Theta}^{*}\left(\kappa,\tau\right)\right)\pto0.
\]
However, the convergence of $\sqrt{n}g_{n}\left(\psi\right)$ to a
normal implies that $\left\Vert \sqrt{n}g_{n}\left(\psi\right)\right\Vert ^{-1}$
converges in distribution to a random variable with zero mass at zero,
and thus that (\ref{eq: first term convergence})
holds, as we wished to show.

Our argument thus far implies that for sufficiently large $\kappa\ge1$, and each $\nu>0,$
\[
\liminf_{n\to\infty}\P\left(\sqrt{n}\cdot d_{H}\left(\hat{\Theta}\left(\kappa,\tau\right),\hat{\Theta}(\tau)\right)\le d_{H}\left(\hat{\Theta}^{*}\left(\kappa,\tau\right),\hat{\Theta}^{*}\left(\tau\right)\right)+\nu\right)=1.
\]
To complete the proof, we wish to show that 
$
d_{H}\left(\hat{\Theta}^{*}\left(\kappa,\tau\right),\hat{\Theta}^{*}\left(\tau\right)\right)\pto 0
$
as $\kappa\to\infty$. To this end, note that $\hat{\Theta}^{*}\left(\kappa,\tau\right)$
and $\hat{\Theta}^{*}\left(\tau\right)$ both depend on the data only
through $\sqrt{n}g_{n}(\psi)$, and let 
\[
\mathcal{G}(\kappa,\zeta)=\left\{ \sqrt{n}g_{n}(\psi):d_{H}\left(\hat{\Theta}^{*}\left(\kappa,\tau\right),\hat{\Theta}^{*}\left(\tau\right)\right)<\zeta\right\} .
\]
Note that since $\mathcal{W}_{\kappa}$ is a compact, connected set and $h'\hat{\phi}_{\Omega}^{*}$
is continuous in $\Omega$, $\hat{\Theta}^{*}\left(\kappa,\tau\right)$
is a closed interval. Since $\hat{\Theta}^{*}\left(\tau\right)$ is
likewise an interval, $d_{H}\left(\hat{\Theta}^{*}\left(\kappa,\tau\right),\hat{\Theta}^{*}\left(\tau\right)\right)$
is simply the maximum of the distance between the upper and lower
endpoints. Moreover, since the endpoints of both intervals vary continuously
in $\sqrt{n}g_{n}(\psi)$, $d_{H}\left(\hat{\Theta}^{*}\left(\kappa,\tau\right),\hat{\Theta}^{*}\left(\tau\right)\right)$
is also continuous in $\sqrt{n}g_{n}(\psi),$ so $\mathcal{G}(\kappa,\zeta)$
is an open set. By the Portmanteau lemma and convergence in distribution
of $\sqrt{n}g_{n}(\psi),$ it follows that 
\[
\liminf_{n\to\infty}\P\left\{
d_{H}\left(\hat{\Theta}^{*}\left(\kappa,\tau\right),\hat{\Theta}^{*}\left(\tau\right)\right)<\zeta\right\}
\ge\P\left\{ \Norm \left(\eta,\Sigma\right)\in\mathcal{G}(\kappa,\zeta)\right\} ,
\]
and thus that the asymptotic probability that $d_{H}\left(\hat{\Theta}^{*}\left(\kappa,\tau\right),\hat{\Theta}^{*}\left(\tau\right)\right)<\zeta$
is lower bounded by the analogous probability in the normal limit
problem. The proof of \cref{prop:J-stat} implies that for
each fixed $\sqrt{n}g_{n}\left(\psi\right),$ however, $d_{H}\left(\hat{\Theta}^{*}\left(\kappa,\tau\right),\hat{\Theta}^{*}\left(\tau\right)\right)=0$
for $\kappa$ sufficiently large, so 
\[
\lim_{\kappa\to\infty}\P\left\{ N\left(\eta,\Sigma\right)\in\mathcal{G}
(\kappa,\zeta)\right\} =1,
\]
which implies the result.

\end{proof}

\cortstatmainasymptotic*

\begin{proof}

Note that 
\[
t_{\Omega}(\theta_0) = \frac{\hat\theta_\Omega-\theta_0}{\hat\sigma_{\Omega}}=\frac{\sqrt{n}(\hat\theta_\Omega-\vartheta(\psi))}{\sqrt{n}\hat\sigma_{\Omega}}+\frac{\sqrt{n}(\vartheta(\psi)-\theta_0)}{\sqrt{n}\hat\sigma_{\Omega}}
\]
Hence, by \eqref{eq:estimator and se convergence}, for any $\kappa\ge1$
\[
t_{\Omega}(\theta_0)\to_p t^*_\Omega(\theta_0)=\frac{\hat\theta^*_\Omega}{\sigma_\Omega}+\frac{\vartheta(\psi)-\theta_0}{\sigma_\Omega}
\]
uniformly over $\Omega\in\mathcal{W}_\kappa$ and $\theta_0\in[\vartheta(\psi)\pm \frac{1}{\sqrt{n}}C]$ for any fixed $C$.  This implies, in particular, that 
\begin{equation}\label{eq: t convergence}
\inf_{\theta_0\in [\vartheta(\psi)\pm \frac{1}{\sqrt{n}}C]}\sup_{\Omega\in\mathcal{W}_\kappa}\left|t_\Omega(\theta_0)\right|\pto
\inf_{\theta_0\in [\vartheta(\psi)\pm \frac{1}{\sqrt{n}}C]}\sup_{\Omega\in\mathcal{W}_\kappa}\left|t^*_\Omega(\theta_0)\right|.
\end{equation}
Note that by the proof of \cref{cor:tstatmain}, $\left|t^*_\Omega\left(\vartheta(\psi)+\frac{1}{\sqrt{n}}\hat\theta^*_{\Sigma^{-1}}\right)\right|\le\sqrt{J^*}$ for all $\Omega.$  Hence, for any $C\ge\left|\hat\theta^*_{\Sigma^{-1}}\right|,$ the right-hand side or \eqref{eq: t convergence} is bounded above by $\sqrt{J^*,}$ regardless of $\kappa.$  Since $\sqrt{n}g_n(\psi)$ converges in distribution to a tight limit, so does $\hat\theta^*_{\Sigma^{-1}},$ so since the left hand side of \eqref{eq: t convergence} is decreasing in $C,$ and $\sqrt{J}-\sqrt{J^*}\pto 0$ as shown in the proof of \cref{prop:J-stat-asymptotic}, we have that
\begin{equation}\label{eq: asymptotic lower bound}
\limsup_{n\to\infty}\P\left\{\inf_{\theta_0}\sup_{\Omega\in\mathcal{W}_k}|t_\Omega(\theta_0)|>\sqrt{J}+\varepsilon\right\}=0
\end{equation}
for all $\kappa\ge1$, $\varepsilon>0.$

Conversely, note that for any $\tau>0$ and any $\varepsilon>0,$ \cref{prop:J-stat-asymptotic} implies that there exist $\kappa$ and (random) weight matrices $\Omega_+,\Omega_-\in\mathcal{W}_\kappa$ such that $\hat\sigma_{\Omega_+},$ $\hat\sigma_{\Omega_-}\le\sqrt{1+\tau^2}\hat\sigma_{\Sigma^{-1}}$ and $\Omega_+,\Omega_-$ approximately obtain the bounds in the proposition,
\begin{equation}
\begin{split}
\label{eq:omega+-def}
&\limsup_{n\to\infty}\P\left\{\sqrt{n}\left|\hat\theta_{\Omega_+}-\hat\theta_{\Sigma^{-1}}-\sqrt{J}\hat\sigma_{\Sigma^{-1}}\tau\right|>\varepsilon\right\}<\varepsilon, \\
&\limsup_{n\to\infty}\P\left\{\sqrt{n}\left|\hat\theta_{\Omega_-}-\hat\theta_{\Sigma^{-1}}+\sqrt{J}\hat\sigma_{\Sigma^{-1}}\tau\right|>\varepsilon\right\}<\varepsilon.
\end{split}
\end{equation}
Since this holds for all $\varepsilon>0,$ for all $\varepsilon'>0$ there exists $\kappa$ and $\Omega_+,\Omega_-\in\mathcal{W}_\kappa$ such that 
\begin{equation}
\begin{split}
\label{eq:tstat+-bound}
   &\limsup_{n\to\infty}\P\left\{\left|t_{\Omega_+}(\hat\theta_{\Sigma^{-1}})\right|\le \frac{\tau}{\sqrt{1+\tau^2}}\sqrt{J}-\varepsilon'\right\}<\varepsilon', \\
   &\limsup_{n\to\infty}\P\left\{\left|t_{\Omega_-}(\hat\theta_{\Sigma^{-1}})\right|\le \frac{\tau}{\sqrt{1+\tau^2}}\sqrt{J}-\varepsilon'\right\}<\varepsilon'.
\end{split}
\end{equation}
Note, however, that for all $\theta_0,$ 
\[
\sup_{\Omega\in\mathcal{W}_\kappa}|t_{\Omega}(\theta_0)|\ge\max\left\{\left|t_{\Omega_+}(\theta_0)\right|,\left|t_{\Omega_-}(\theta_0)\right|\right\}.
\]
However, since (with high probability) $|t_{\Omega_+}(\theta_0)|$ is decreasing in $\theta_0$ for $\theta_0\le\hat\theta_{\Sigma^{-1}}$ while $|t_{\Omega_-}(\theta_0)|$ is increasing in $\theta_0$ for $\theta_0\ge\theta_{\Sigma^{-1}},$
\[
\max\left\{\left|t_{\Omega_+}(\theta_0)\right|,\left|t_{\Omega_-}(\theta_0)\right|\right\}\ge\min\left\{\left|t_{\Omega_+}(\hat\theta_{\Sigma^{-1}})\right|,\left|t_{\Omega_-}(\hat\theta_{\Sigma^{-1}}))\right|\right\}.
\]
However, we have shown that for $\kappa$ sufficiently large, with high probability we can lower bound the right hand side by $\sqrt{J}-\varepsilon'$ for any $\varepsilon'>0.$  Thus, for all $\varepsilon>0$ and $\tau>0,$
\[
\lim_{\kappa\to\infty}\limsup_{n\to\infty}\P\left\{\inf_{\theta_0}\sup_{\Omega\in\mathcal{W}_{\kappa}}\left|t_\Omega(\theta_0)\right|<\frac{\tau}{\sqrt{1+\tau^2}}\sqrt{J}-\varepsilon\right\}=0.
\]
Together with \eqref{eq: asymptotic lower bound}, this implies the result.

\end{proof}

\corcsasymptotic*

\begin{proof}

The first part of the result is immediate from \cref{cor:tstatmain-asymptotic}.  For the second part, note that the proof of
\cref{cor:tstatmain-asymptotic} shows that $\left|t^*_\Omega\left(\vartheta(\psi)+\frac{1}{\sqrt{n}}\hat\theta^*_{\Sigma^{-1}}\right)\right|\le\sqrt{J^*}$ for all $\Omega,$ while by \eqref{eq:estimator and se convergence},
$t^*_\Omega\left(\vartheta(\psi)+\frac{1}{\sqrt{n}}\hat\theta^*_{\Sigma^{-1}}\right)- t_\Omega(\hat\theta_{\Sigma^{-1}})\pto 0$ uniformly over $\Omega\in\mathcal{W}_\kappa$ and $\sqrt{J^*}-\sqrt{J}\pto0.$  Thus, for all $\varepsilon'$
\[
\limsup_{n\to\infty}\P\left\{\sup_{\Omega\in\mathcal{W}_\kappa}\left|t_\Omega\left(\hat\theta_{\Sigma^{-1}}\right)-\sqrt{J}\right|>\varepsilon'\right\}\to 0,
\]
and $\hat\theta_{\Sigma^{-1}}\in \bigcap_{\Omega\in\mathcal{W}_{\kappa}} C_\Omega(\sqrt{J}+\nu)$ with probability going to one for any $\nu>0$ and any $\kappa.$

On the other hand, note that since $C_\Omega(c)$ is an interval for all $c,$ $\bigcap_{\Omega\in\mathcal{W}_{\kappa}} C_\Omega(\sqrt{J}+\nu)$ is also an interval.  For any $\tau>0$ and any $\nu'>0$, following the proof of \cref{cor:tstatmain-asymptotic}, there exists $\kappa\ge1$ and (random) $\Omega_+$ and $\Omega_-\in\mathcal{W}_\kappa$ such that $\hat\sigma_{\Omega_+},$ $\hat\sigma_{\Omega_-}\le\sqrt{1+\tau^2}\hat\sigma_{\Sigma^{-1}}$, and \eqref{eq:omega+-def} and \eqref{eq:tstat+-bound} hold (with $\nu'$ in place of $\varepsilon$ and $\varepsilon'$).
As also argued in the proof of \cref{cor:tstatmain-asymptotic}, for all $\theta_0$,
\[
\sup_{\Omega\in\mathcal{W}_\kappa}|t_{\Omega}(\theta_0)|\ge\min\left\{\left|t_{\Omega_+}(\theta_0)\right|,\left|t_{\Omega_-}(\theta_0)\right|\right\}.
\]
Note, moreover, that $|t_{\Omega_+}(\theta_0)|$ is decreasing in $\theta_0$ for $\theta_0\le \hat\theta_{\Sigma^{-1}},$ with slope $1/\hat\sigma_{\Omega_+},$
while $|t_{\Omega_-}(\theta_0)|$ is increasing in $\theta_0$ for $\theta_0\ge \hat\theta_{\Sigma^{-1}},$ with slope $1/\hat\sigma_{\Omega_-}$.  Consequently, for all $\theta_0$
\[
\sup_{\Omega\in\mathcal{W}_\kappa}|t_{\Omega}(\theta_0)|\ge \min\left\{t_{\Omega_+}\left(\hat\theta_{\Sigma^{-1}}\right),t_{\Omega_-}\left(\hat\theta_{\Sigma^{-1}}\right)\right\}+\frac{\left|\hat\theta_{\Sigma^{-1}}-\theta_0\right|}{\max\left\{\hat\sigma_{\Omega_+},\hat\sigma_{\Omega_-}\right\}}\ge
\]
\[ \min\left\{t_{\Omega_+}\left(\hat\theta_{\Sigma^{-1}}\right),t_{\Omega_-}\left(\hat\theta_{\Sigma^{-1}}\right)\right\}+\frac{\left|\hat\theta_{\Sigma^{-1}}-\theta_0\right|}{\sqrt{1+\tau^2}\hat\sigma_{\Sigma^{-1}}}.
\]
Consequently, for any $\delta,$
\[
\limsup_{n\to\infty}\P\left\{
\sup_{\Omega\in\mathcal{W}_\kappa}\left|t_{\Omega}\left(\hat\theta_{\Sigma^{-1}}+\frac{1}{\sqrt{n}}\delta\right)
\right|\le \frac{\tau}{\sqrt{1+\tau^2}}\sqrt{J}+\frac{|\delta|}{\sqrt{1+\tau^2}\sigma_{\Sigma^{-1}}}-\nu'\right\}<2\nu'.
\]
Since we can repeat this argument for all $\nu'>0$ by taking $\kappa$ sufficiently large, it follows that for all $\nu'$ and $\tau>0,$
\[
\lim_{\kappa\to\infty}\limsup_{n\to\infty}\P\left\{\sup_{\Omega\in\mathcal{W}_\kappa}\left|t_{\Omega}\left(\hat\theta_{\Sigma^{-1}}+\frac{1}{\sqrt{n}}\delta\right)
\right|>\frac{\tau}{\sqrt{1+\tau^2}}\sqrt{J}+\frac{|\delta|}{\sqrt{1+\tau^2}\sigma_
{\Sigma^ {-1}}}-\nu'\right\}=1.
\]
Since $J$ converges in distribution to a tight random variable, however, for all $\delta\neq 0$
\[
\lim_{\nu\to0}\sup_{\tau>0}\liminf_{n\to\infty}\P\left\{\sqrt{J}+\nu<\frac{\tau}{
\sqrt{1+\tau^2}}\sqrt{J}+\frac{|\delta|}{\sqrt{1+\tau^2}\sigma_{\Sigma^{-1}}} \right\}=1,
\]
and consequently
\[
\lim_{\nu\to0}\lim_{\kappa\to\infty}\limsup_{n\to\infty}\P\left\{\sup_{\Omega\in\mathcal{W}_\kappa}\left|t_{\Omega}\left(\hat\theta_{\Sigma^{-1}}+\frac{1}{\sqrt{n}}\delta\right)
\right|>\sqrt{J}+\nu\right\}=1.
\]
Since this holds for all $\delta\neq0$ and $\bigcap_{\Omega\in\mathcal{W}_{\kappa}} C_\Omega(\sqrt{J}+\nu)$ is an interval, the second part of the result is immediate.
\end{proof}

\clearpage

\end{document}

%% file: tabJstats.tex
\begin{tabular}{l l c c c l}
\hline\hline\\[-0.9em]
Paper & Model & $\sqrt{J}$ & dF & p-value & Reported in the paper? \\[0.1em]
\hline\hline\\[-0.8em]
 & Table 4, Column 1 & 1.75  & 1 & 0.08 & \\[0.2em]
     \cite{abebe2021selectionoftalent}                             & Table 4, Column 2 & 2.16  & 3 & 0.20 & No \\[0.2em]
                                  & Table 4, Column 4 & 2.07  & 2 & 0.12 & \\[1em]
\cite{bourreau2021marketentry}    & Table 4, Column 2 & 7.13 & 7 & 0.00 & J-statistic and dF \\[1em]
\cite{mueller2021jobseekers}      & Table 5           & 0.68 & 3 & 0.93 & No \\[1em]
\cite{gj2022information}          & Table 2, Column 2 & 1.47  & 5 & 0.83 & J-statistic and p-value \\[1em]
     & Table 6, Column 7 & 0.60 & 1 & 0.55 & \\[0.2em]
        \cite{karahan2024demographic}                          & Table 7, Column 4 & 2.41 & 2 & 0.05 & p-value \\[0.2em]
                                  & Table 7, Column 5 & 0.57 & 1 & 0.57 & \\[1em]
\cite{kelley2024monitoring}       & Table 7           & 0.07 & 1 & 0.94 & No \\[0.2em] 
\hline\hline
\end{tabular}